# Laser-Enhanced Contact Optimization in Silicon Photovoltaics: Mechanisms, Reliability, and Predictive Process Design

**Donald Intal[a,*], Abasifreke U. Ebong[a]**

[a]Department of Electrical and Computer Engineering, University of North Carolina at Charlotte, 9201 University Blvd, Charlotte, 28223, North Carolina, USA

[*]Correspondence: dintal@charlotte.edu (D. Intal); aebong1@charlotte.edu (A.U. Ebong)

ORCID(s): 0000-0003-3528-4894 (D. Intal); 0000-0002-8808-5461 (A.U. Ebong)

**Abstract:** Laser-enhanced contact optimization (LECO) has emerged as an important method for simultaneously reducing contact resistivity and metallization-induced recombination in advanced crystalline silicon solar cells, thereby enabling concurrent gains in fill factor and open-circuit voltage, particularly in TOPCon devices. However, broader industrial transferability remains constrained by the need to preserve these gains within a narrow process window and by unresolved, architecture-dependent questions regarding the kinetic stability of some LECO-modified interfaces. LECO is therefore examined in this review as a coupled multiphysics process that links localized electrothermal activation and microstructural evolution to device-level electrical signatures through an instantaneous regime map and a reliability classification based on time-dependent drift. A predictive workflow is outlined that couples transient electrothermal modeling with reduced state metrics, including effective diffusion depth and local areal energy density, and propagates calibrated thresholds across the recipe space. The framework separates stable optimization from marginal activation and latent damage, while explaining why fine-line scaling and copper-containing contact stacks can tighten stability margins through current localization and diffusion-barrier constraints. These insights provide a basis for reliability-aware process-window design and future digital-twin-assisted optimization of LECO for scalable, high-efficiency silicon photovoltaics.

---

# 1. Introduction

The steady increase in conversion efficiency of crystalline silicon photovoltaics has been driven by systematic optimization of device architecture having low surface and metal recombination losses. The passivated emitter rear (PERC) and tunnel oxide passivated contact (TOPCon) solar cell architectures employ lightly doped, high sheet resistance emitters to suppress surface recombination for enhanced open circuit voltage $V_{OC}$ [1]. In the TOPCon architecture, which now exceeds 25% certified efficiency at industrial scale [2], boron diffused front emitter with sheet resistance approaching 400 to 420 Ω/□ have become a standard. Reducing the surface carrier concentration (i.e., increasing the emitter sheet resistance) lowers the emitter saturation current density $j_{0,e}$, while the selective emitter concept, which allows the contact regions to be heavily doped decreases the metal recombination density $j_{0,met}$. However, with homogeneous emitter, the high sheet resistance increases the specific contact resistivity $\rho_c$, thereby limiting the achievable fill factor [1,3]. The resulting rise in series resistance degrades fill factor, particularly in conjunction with fine line metallization. This coupling is illustrated in Figure 1 in conjunction with the transfer width equation (1) governed by the lateral current crowding beneath the contact fingers [4].

$$W_T = \sqrt{\frac{\rho_c}{R_{sheet}}} \tag{1}$$

As sheet resistance increases, even with modest increase in $\rho_c$, the current constriction and local resistive loss, significantly increase. Consequently, further gains in emitter passivation shift the dominant performance constraint from bulk recombination to the metal semiconductor interface [3].

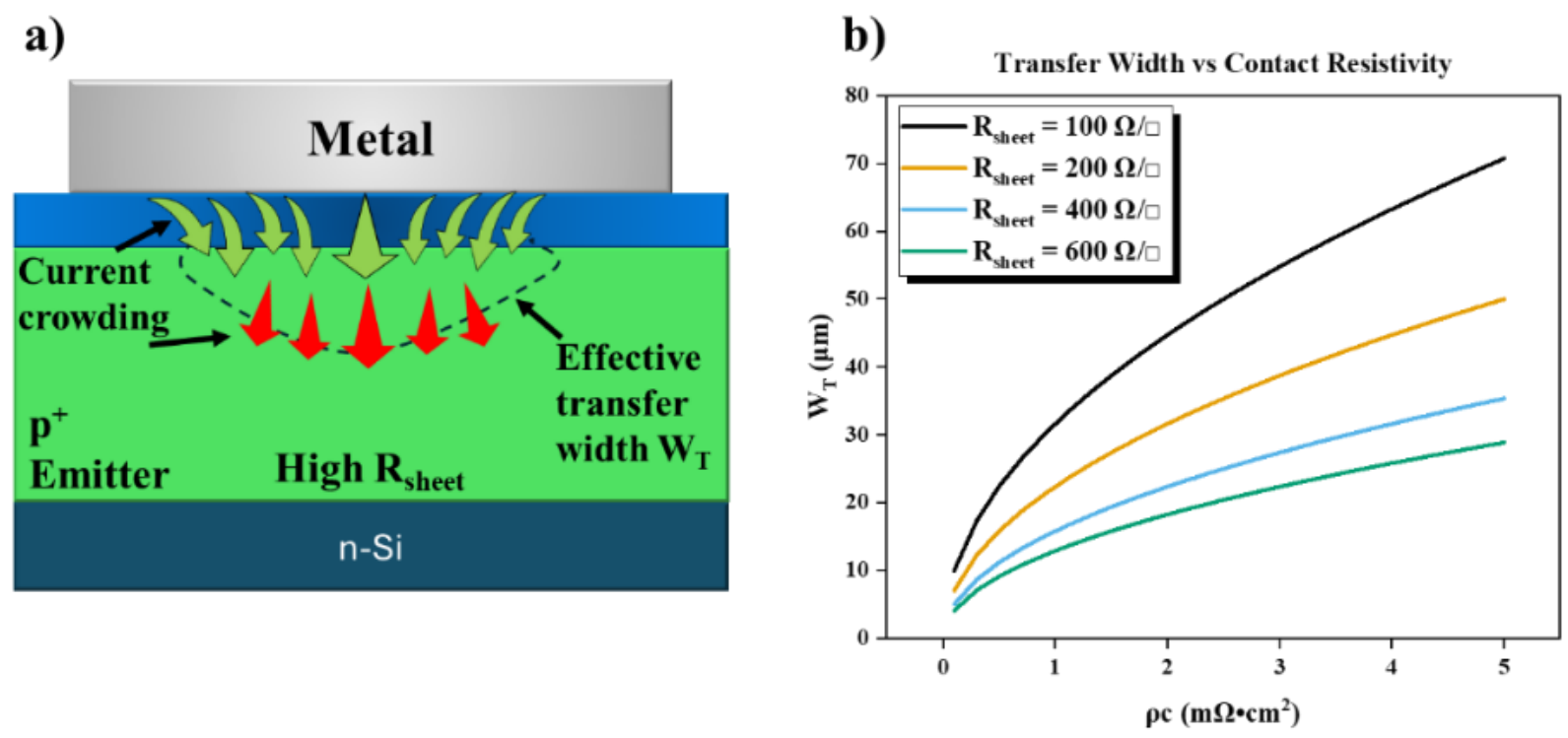

*Figure 1: Transport constraints in high $R_{sheet}$ emitter. (a) Schematic of lateral current flow and vertical injection beneath a metal contact, defining the transfer width $W_T$. (b) Calculated $W_T = \sqrt{\frac{\rho_c}{R_{set}}}$ versus specific contact resistivity $\rho_c$ for representative sheet resistance, illustrating increased sensitivity to $\rho_c$ at higher $R_{sheet}$.*

Conventional screen-printed silver metallization is increasingly challenged by this shift [4]. Achieving low contact resistivity on high $R_{\text{sheet}}$ boron emitters requires aggressive glass frit chemistries to promote Ag-Si interdiffusion during firing [4,5]. However, excessive reactivity risks dopant depletion, deep alloy penetration, and junction damage [5]. The addition of aluminum to silver pastes facilitates the formation of Ag-Al wedge spikes that extend beyond 1 μm and provide conductive pathways through high resistance interfaces [5]. Although effective in lowering $\rho_c$, these structures increase contact recombination and introduce reliability concerns, particularly under chemically induced corrosion [6,7].

The transition toward narrower fingers further narrows the process window. Reduced contact width lowers optical shading but increases local current density at the interface, intensifying sensitivity to both contact resistivity and defect generation [4,8]. These limitations indicate that conventional fire-through silver metallization is operating near its practical boundary for high sheet resistance emitters [7].

Laser-Enhanced Contact Optimization (LECO) was introduced as a process level response to this constraint [9]. By applying high intensity laser scanning perpendicular to the silver grid under reverse bias, LECO generates localized high current densities at the Ag-Si interface after firing [9,10]. This localized activation promotes controlled silver crystallite growth and enhanced interdiffusion without requiring aggressive paste chemistry or elevated peak firing temperatures [11].

Industrial implementation has demonstrated that LECO can simultaneously reduce specific contact resistivity and metallization-induced recombination, with contact resistivities below 1 mΩ-cm$^2$ and $j_{0,\text{met}}$ values below 160 fA/cm$^2$ reported for industrial TOPCon cells. This combination enables concurrent improvements in fill factor and open-circuit voltage. Industrial Q.ANTUM NEO cells have reported $V_{OC}$ above 730 mV after LECO, while recent TOPCon reviews attribute the narrowing voltage gap with silicon heterojunction cells partly to improved contacted passivation and laser-assisted firing technologies [12,13]. Efficiency gains on the order of 0.4 to 0.6% absolute have been reported in large-scale production [14]. LECO therefore relocates the metallization constraint from globally aggressive firing chemistry to localized interfacial transport activation [15].

However, by modifying the interfacial defect landscape and carrier transport pathway, LECO also introduces a new stability axis [16]. Under combined electrical bias and elevated temperature stress, contact resistance at the $p^+$ Ag interface has been observed to increase dramatically in some studies, accompanied by severe efficiency degradation under accelerated conditions [17–19]. This behavior has been associated with hydrogen accumulation at the Ag-Si interface originating from passivation layers such as $AlO_x$ and $SiN_x$ [20,21].

In conventional fire-through contacts, a higher density of unpassivated defects can act as hydrogen sinks [21]. In contrast, the highly passivated interface produced by LECO may alter hydrogen trapping and redistribution dynamics [16,20]. The mechanistic origin, timescale relevance, and field applicability of this degradation remain incompletely resolved [18,19]. At present, direct evidence for LECO-specific long-term instability remains limited and architecture dependent: extended module testing of LECO-treated PERC cells has not revealed abnormal degradation, whereas recent accelerated-stress work on TOPCon has identified a possible bias-thermal failure mode. The central question is therefore how to distinguish robust contact activation from interfacial states that may become unstable under prolonged electrical and thermal stress.

Simultaneously, economic and supply pressures are driving the industry to explore alternatives to silver metallization [7,22]. Screen printed TOPCon cells typically consume between 9 and 15 mg/W of silver, and roadmaps target reductions below 2 mg/W through hybrid Ag-Cu pastes or full copper electroplating [22]. Copper metallized silicon heterojunction devices have demonstrated efficiencies within approximately 0.4 percent absolute of silver references, confirming technical feasibility at the cell level [23–25].

Copper introduces distinct interfacial chemistry, including different diffusion kinetics, oxidation susceptibility, and potential contamination pathways [6,24]. These differences may alter the balance between contact resistivity and recombination relative to silver systems [26]. A critical open question is whether copper-based contacts avoid the transport stability tradeoffs observed in LECO activated silver systems or whether similar hydrogen related or diffusion driven instabilities emerge under long term operation [6]. The material transition may therefore redistribute rather than eliminate the interfacial constraint.

As silicon architectures approach practical efficiency ceilings near 28 percent, incremental performance gains increasingly depend on resolving interfacial transport and long term stability constraints rather than bulk absorber limitations [23,27]. Metallization has therefore emerged as a system level determinant of both manufacturing viability and field durability. The framework presented here provides a basis for evaluating silver and copper contact strategies under a common set of physical principles, with the goal of enabling high initial efficiency without compromising 25-year reliability.

This review develops a unified electrical and mechanistic framework to analyze LECO as a coupled contact-engineering problem in which transport, recombination, architecture, and long-term stability must be optimized together. The central thesis is that LECO performance cannot be evaluated from initial contact resistivity alone; it must be interpreted through coupled transport-recombination gains, microcontact connectivity, and stress-dependent stability margins. Instead of treating contact resistance, recombination, material selection, and reliability as isolated metrics, total series resistance is decomposed into contributions from emitter sheet resistance, contact geometry and transfer width, and interfacial barrier properties. Recombination losses are evaluated through $j_{0,\mathrm{met}}$ and their impact on implied $V_{\mathrm{oc}}$. By integrating transport, recombination, and stability within a single analytical structure, the review clarifies the reliability-performance trade space governing next generation metallization.

# 2. The Physics of Laser Metal Semiconductor Interaction

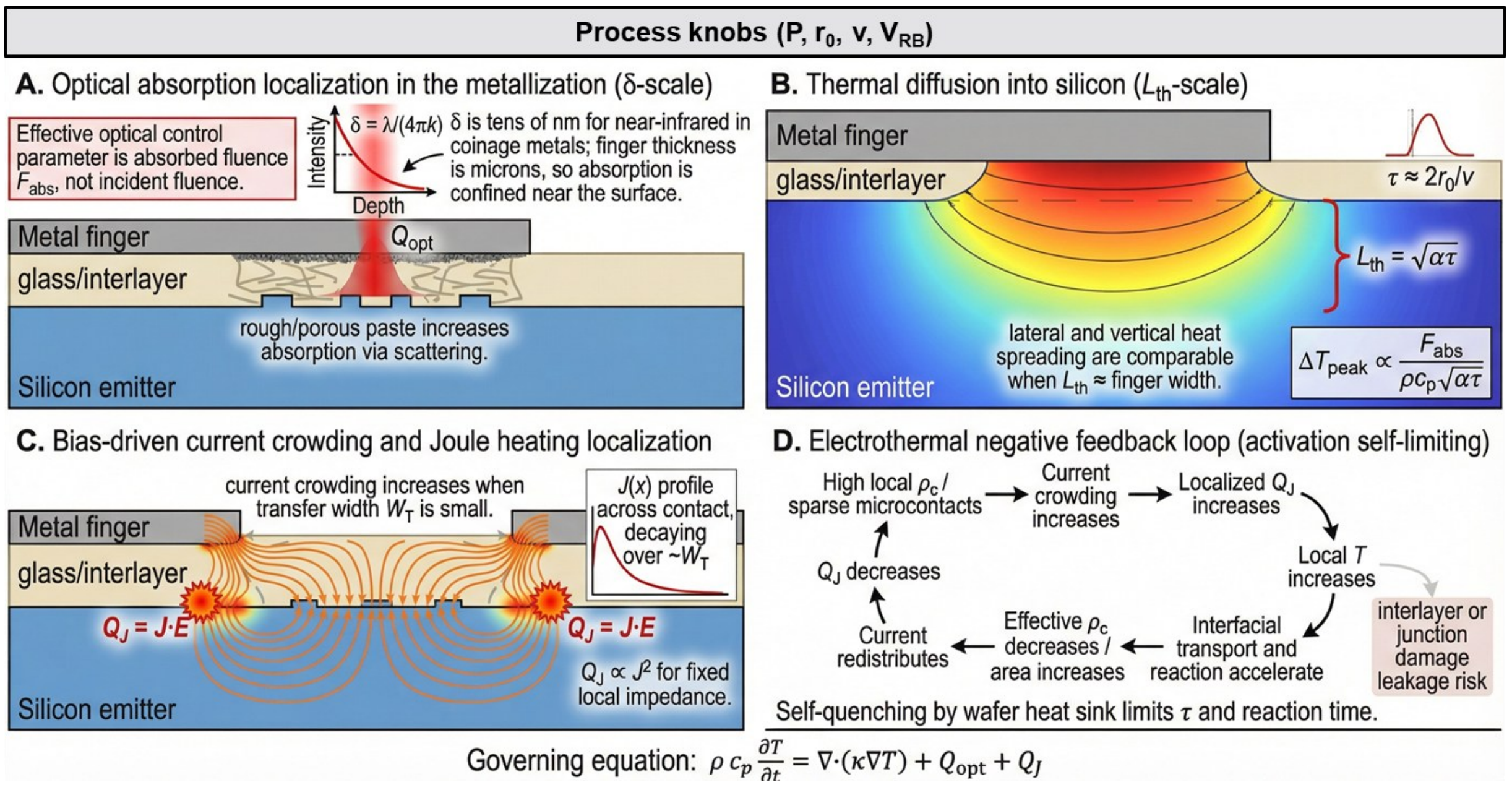

*Figure 2: Coupled optical, thermal, and electrical localization in laser-enhanced contact optimization. (A) Near-surface absorption in the metallization characterized by the optical penetration depth δ. (B) Transient heat diffusion into silicon characterized by $L_{\mathrm{th}} = \sqrt{\alpha\tau}$ and the scaling of $\Delta T_{\mathrm{peak}}$ with absorbed fluence and dwell time. (C) Reverse-bias current crowding over the transfer width $W_T$ produces localized Joule heating $Q_J = J \cdot E$. (D) Electrothermal negative feedback redistributes current as $\rho_c$ decreases, bounding activation within a narrow process window [3,4,9,10,27–30].*

The efficiency gains enabled by LECO can be traced to a short-lived, spatially confined perturbation of the metal-semiconductor contact that couples three processes: (i) optical energy deposition in the front metallization, (ii) transient heat flow into the near-surface silicon, and (iii) bias-driven current transport through an interface whose effective transport properties evolve during the event. A physics-based description is useful because it links LECO control variables (laser power, spot size, scan speed, and applied bias) to the same quantities that govern contact losses in high sheet resistance emitters, namely (i) transfer width, (ii) current crowding, and (iii) interfacial dissipation. Thus, LECO is not a second firing step rather a localized activation event that modifies the interfacial transport pathway while leaving the bulk junction and rear structure largely unaffected. The coupled localization mechanisms, and their connection to the controllable process knobs, are summarized schematically in Figure 2.

A. Optical absorption sets the initial localization of energy. In typical implementations, the laser illuminates the front grid normal to the wafer surface. For a conductor, the relevant absorption length scale can be expressed using the optical penetration depth given in equation (2)

$$\delta = \frac{\lambda}{4\pi k} \tag{2}$$

where k is the extinction coefficient [28]. For near-infrared wavelengths and coinage metals, $\delta$ is on the order of tens of nanometers [29], which is negligible compared with micrometer-scale screen-printed fingers. Consequently, laser energy is deposited in a thin near-surface region of the metallization, producing a steep initial temperature gradient in the metal [29]. The absorbed fraction is not determined solely by bulk optical constants because fired paste surfaces are rough and porous and may contain glass and microstructural texture that reduce specular reflection and increase absorption through scattering and multiple internal reflections [31]. For modeling and scaling, the primary optical control parameter is therefore the absorbed fluence at the finger surface, rather than the incident fluence alone [30].

B. Following absorption, the thermal field evolves through rapid diffusion from the metallization into silicon. A compact descriptor of the heated volume is the thermal diffusion length given in equation (3)

$$L_{\mathrm{th}} = \sqrt{\alpha\tau}, \tag{3}$$

where α is thermal diffusivity and τ is an effective dwell time defined by pulse duration or by spot residence time during scanning [28]. For typical LECO dwell times, $L_{\mathrm{th}}$ is comparable to modern fine-line finger widths, so vertical and lateral heat spreading occur on similar length scales beneath the contact [29]. This geometric coupling implies that the peak interface temperature depends on finger width and local metallization geometry, since lateral leakage into adjacent passivated regions reduces the temperature rise directly under the finger [29]. It also implies that the event is inherently self-quenching because the wafer acts as a large heat sink. Once the spot passes, the near-surface region cools rapidly and limit the time available for diffusion and interfacial reactions [30].

A useful scaling estimate for the characteristic temperature rise under transient surface heating is obtained from diffusion into a semi-infinite medium as in equation (4),

$$\Delta T_{\mathrm{peak}} \sim \frac{2F_{\mathrm{abs}}}{\rho c_p \sqrt{\pi \alpha \tau}}, \tag{4}$$

where $F_{\mathrm{abs}}$ is absorbed fluence and ρ, $c_p$, and α represent effective thermophysical parameters of the region that controls heat removal, often dominated by silicon once heat crosses the thin absorbing metal layer [28,30]. This expression highlights why scan speed and spot size are as important as laser power. They set τ and the spatial distribution of $F_{\mathrm{abs}}$ [29,31]. It also clarifies how LECO accesses interfacial activation without a large global thermal budget. Heating is localized by absorption in the metallization, and subsequent cooling is enforced by diffusion into the wafer [32].

C. The distinguishing feature of LECO relative to laser irradiation alone is that the thermal transient occurs under an applied electrical bias [3]. As illustrated in Figure 2C,D, bias-driven current flow introduces an additional dissipation channel at the interface and modifies the spatial distribution of heating through current crowding [4,10]. At the continuum level, the coupled electrothermal problem can be written as in equation (5)

$$\rho c_p \frac{\partial T}{\partial t} = \nabla \cdot (k \nabla T) + Q_{\mathrm{opt}} + Q_{\mathrm{J}}, \tag{5}$$

where $Q_{\mathrm{opt}}$ represents optical absorption in the metallization and $Q_{\mathrm{J}}$ represents Joule heating associated with current flow through resistive pathways[29,30,32], locally as in equation (6)

$$Q_{\mathrm{J}} = J \cdot E. \tag{6}$$

Because the contact is initially spatially inhomogeneous and can exhibit high effective $\rho_c$ after firing, current preferentially enters through the lowest-resistance micro-paths, producing strong current crowding near the contact perimeter [4,6]. The relevance of this effect is directly connected to the transfer width framework introduced in Sec. 1. When $W_T$ is small, the injection region is narrow and local J can greatly exceed its area-averaged value [27]. Since $Q_{\mathrm{J}}$ scales with $J^2$ for a given local impedance, Joule heating becomes strongly localized at precisely the regions that limit transport [10].

D. This coupling provides a natural feedback mechanism for contact activation. Regions that initially limit transport experience the largest $Q_{\mathrm{J}}$, which raises the local temperature and accelerates processes that reduce the effective barrier to transport and increase the conductive interfacial area [3,9]. As the effective contact improves and the local resistivity decreases, current redistributes over a larger area, reducing crowding and lowering $Q_{\mathrm{J}}$, which limits further modification [9]. This electrothermal negative feedback explains, at a mechanistic level, why LECO can reduce effective contact resistance without requiring globally aggressive firing conditions and the optimization requires both optical and electrical parameters [3].

Beyond Joule heating, the applied field can in principle influence mass transport when mobile species are present at elevated temperature. A generic drift-diffusion form for a mobile metal species flux is shown in equation (7)

$$J_{\mathrm{M}} = -D\frac{\partial C}{\partial x} + \frac{DCZ^{*}eE}{k_B T}, \quad (7)$$

where the second term represents field-driven drift [33]. Whether this contribution is significant depends on the system and operating regime and is difficult to isolate experimentally because electric field, current density, and temperature evolve together as the interface changes [16]. The key point is that, bias in LECO is not merely an external boundary condition. It can alter both the spatial distribution of dissipation and the kinetics of interfacial reconfiguration through the coupled dependence of $\rho_c$, current crowding, and temperature [9,10].

The same localization that enables low initial contact resistance can produce an interfacial state that is structurally and chemically distinct from conventional fire-through contacts, with a different defect population and species distribution in the near-interface region. Under subsequent field and temperature stress, that interfacial state can evolve, potentially increasing contact resistance and degrading fill factor. Section 2 therefore provides the baseline. LECO should be treated as a localized electrothermal activation of transport that is inseparable from current-crowding physics in high $R_{\text{sheet}}$ emitters. Subsequent sections map this baseline onto specific microstructural pathways for $\rho_c$ reduction and onto the mechanisms that control long-term drift in transport and recombination.

The LECO process window is governed by four coupled quantities: absorbed fluence, dwell time, reverse-bias current localization, and the initial heterogeneity of the fired contact. Absorbed fluence determines the magnitude of the thermal impulse delivered to the metallization, while dwell time, set by scan speed and spot size, determines how long the interface remains within the activation range. Reverse-bias current localization determines where Joule heating is concentrated, making LECO self-selective toward initially resistive or current-crowded microcontacts. The initial fired-contact heterogeneity then determines whether activation proceeds through a broad population of weak sites or through a small number of dominant hot spots.

Laser power alone is therefore an incomplete descriptor of LECO intensity. Two recipes with the same nominal laser power can generate different interfacial states if scan speed, spot diameter, finger geometry, or reverse-bias condition changes the local areal energy density and current-crowding factor. A practical process map should therefore use reduced variables such as local areal energy density, maximum interface temperature, effective diffusion depth, and current-crowding factor. The stable process window is reached when these variables are sufficient to increase microcontact connectivity and reduce contact resistivity, but remain below the thresholds for excessive metal penetration, leakage, or recombination-active damage.

# 3. Interfacial Microstructural Evolution

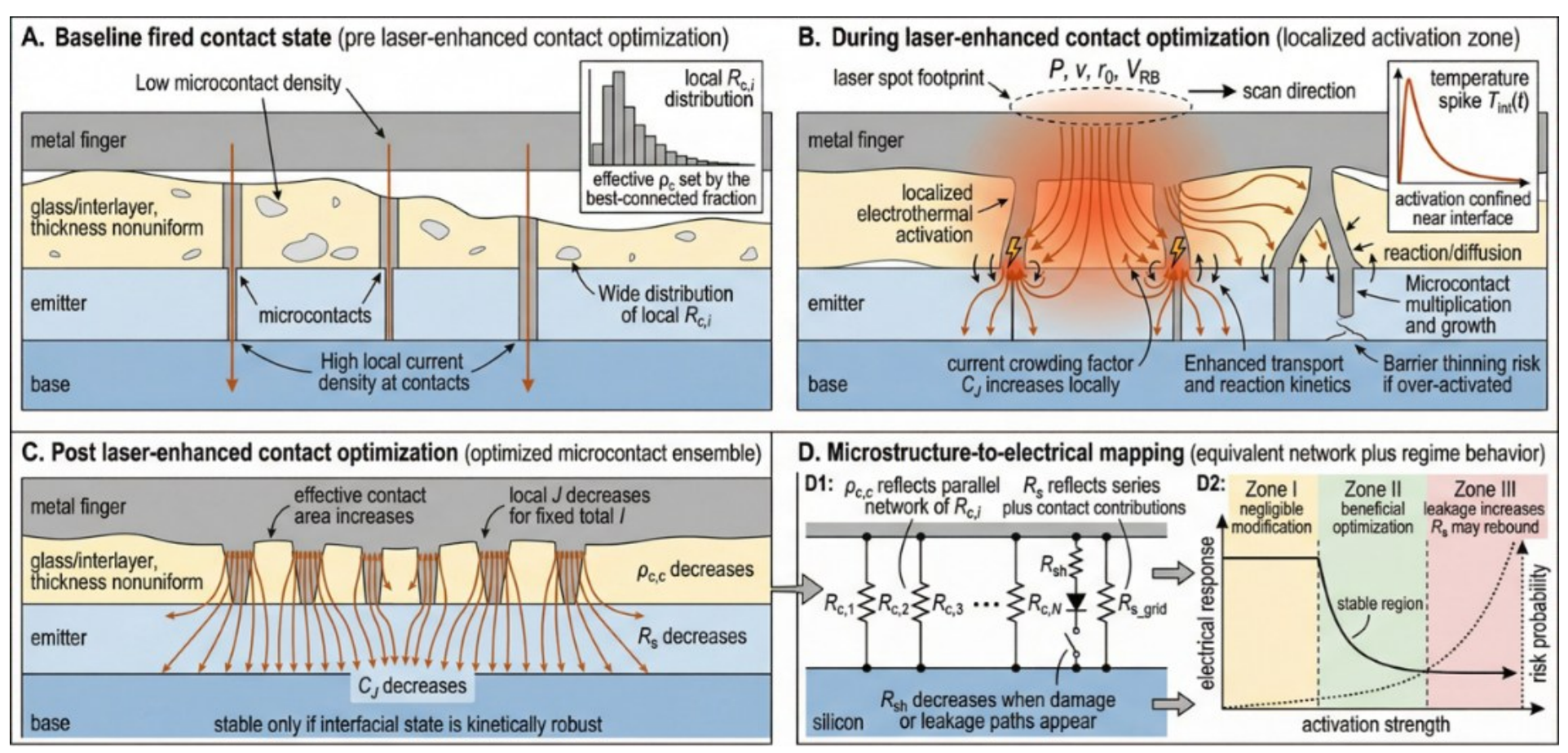

*Figure 3: Laser-enhanced contact optimization as microcontact-ensemble engineering. (A) Baseline contacts are heterogeneous and current localizes through sparse microcontacts. (B) Localized electrothermal activation under reverse bias drives microcontact growth and multiplication, with over-activation risk. (C) Increased microcontact connectivity reduces current crowding and lowers effective* $\rho_c$ *and* $R_s$ *when kinetically stable. (D) Equivalent network and regime schematic link microstructure to electrical response and degradation risk.*

LECO contact modification is local due to optical absorption and bias-driven Joule heating; the interfacial reactions nucleate at discrete hot spots that coincide with the highest-current micro-paths [9,10]. The processed interface therefore departs from an idealized planar contact and is better represented as a heterogeneous ensemble of conductive microcontacts embedded in a more resistive matrix composed of residual glass, oxides, and partially reacted silicon [4,5]. In this regime, the measured specific contact resistivity $\rho_c$ is governed by the statistics and connectivity of the microcontact network rather than by the metallized area alone [4]. This microcontact-ensemble is shown in Figure 3, provides a common scenario for both Ag- and Cu-based contacts. For silver-based contacts on passivated architectures, conductive features are often reported as localized Ag-Si reaction volumes, frequently discussed as current-fired contact structures, that remain confined near the doped poly-Si region [5,10]. For copper-based systems, interfacial evolution is typically mediated by silicide formation, most prominently $Cu_3Si$, which thickness and continuity depend on local thermal budget and barrier integrity [6,26]. These observations motivate a unified interpretation of LECO contacts as disordered transport networks, where abrupt reductions in $\rho_c$ correspond to the emergence of sufficiently connected low-impedance pathways [4,11].

## 3.1 Silver Contacts: Activation and Spike Evolution

In conventional fire-through Ag or Ag/Al metallization, contact formation is mediated by the glass frit. During firing, the glass softens, wets the dielectric (for example $SiN_x$), locally etches toward the silicon stack, and enables Ag-rich protrusions or spikes to develop as the interface becomes chemically and structurally reactive [34–36]. Formulations that promote early wetting and rapid interfacial spreading can reduce $\rho_c$, but excessive glass penetration and alloying increase the probability of junction damage and metallization-induced recombination [35,36]. This trade-off becomes especially restrictive for high $R_{\text{sheet}}$ emitters and fine-line fingers, where the incentive to lower $\rho_c$ is highest and current crowding amplifies local interfacial dissipation [37–40].

LECO modifies this pathway by adding a localized electro-thermal transient that selectively intensifies reaction at current-crowded micro-sites [39,40]. Rather than relying solely on the global firing thermal budget and globally reactive frit chemistry, LECO concentrates energy into discrete locations determined by the paste microstructure, local barrier thickness variations, and the electrical boundary condition imposed during processing [34,41]. A consistent microstructural interpretation is that local temperatures can transiently approach the Ag-Si eutectic regime at specific sites, producing brief liquid-like Ag-Si reaction volumes that resolidify as fine eutectic microstructures [36,38]. Importantly, the resulting interface is typically not described by continuous deep spikes extending far into bulk silicon [37,42]. Instead, microstructural analyses commonly report isolated Ag-Si reaction zones (often described as CFC structures) located within or just above the doped poly-Si region, with the underlying tunnel oxide frequently reported to remain largely intact [37,38]. This confinement is central to the LECO value proposition: it increases conductive coupling into the doped region while reducing the prevalence of long, recombination-active intrusions associated with aggressive fire-through reactions.

From a transport perspective, the microstructural outcome is a redistribution of the true electrical contact area into a population of low-impedance patches [40,42]. LECO increases the number and lateral extent of these patches without converting the entire nominal pad into a continuous metal-semiconductor junction [37,42]. Residual glass or thin oxide remnants persist outside the most conductive zones, implying that carrier injection remains spatially heterogeneous [42,43]. In those surrounding regions, transport plausibly proceeds through tunneling or field-assisted conduction through thin barriers into adjacent doped poly-Si, while eutectic microcontacts act as low-resistance nodes that dominate current collection and injection [35,36,38,42]. This supports a hybrid mechanism: LECO reduces effective $\rho_c$ both by increasing the areal density of low-barrier microcontacts and by locally thinning or displacing resistive interlayers along preferred current paths [37,39,42]. The relative weight of these contributions depends on paste composition, firing history, and the initial barrier landscape, but the defining feature is the same: LECO shifts contact formation toward a discrete microcontact ensemble [37,40].

## 3.2 Copper Contact: Silicide-Mediated Contact Formation

For copper metallization, ohmic coupling to silicon is typically mediated by silicide formation, most commonly $Cu_3Si$ [6,44]. $Cu_3Si$ can nucleate at relatively low temperatures and can grow rapidly once the interface becomes reactive [44]. This creates a narrow design window: a thin, continuous silicide can lower $\rho_c$ and stabilize current injection, whereas uncontrolled growth can consume silicon, generate voiding, and create shunting pathways [6,44,45]. Thin-film and nanolayer studies consistently show an initial intermixing stage followed by $Cu_3Si$ formation and subsequent thickening with time and temperature [44,46]. Under strong thermal gradients or locally elevated peak temperatures, growth can become abrupt and nonuniform, reflecting diffusion asymmetry and transport behavior that can deviate from simple Fickian expectations [45,47]. In a LECO-like transient, the key risk is therefore not the average temperature but the local peak temperature and dwell-time tails at current-crowded sites, which can push silicide evolution from bounded to runaway [45,46].

Industrial copper contact stacks therefore employ barrier layers, often Ni or Ni-based alloys, between Cu and Si to suppress uncontrolled interdiffusion while still enabling controlled formation of an interfacial seed phase [6,48]. In this design space, the controlling variables are (i) barrier thickness and continuity, (ii) the identity of any barrier silicide that forms under thermal exposure, and (iii) the local thermal budget delivered during rapid processing [46,48,49]. In a controlled regime, the barrier remains intact and silicide formation is confined to a thin, continuous layer that improves electrical coupling without compromising passivation [48]. In an uncontrolled regime, barrier discontinuities, local overheating, or excessive cumulative thermal exposure allow Cu to bypass the barrier and form thick, irregular silicide penetration and voided morphologies, correlating with leakage currents, local recombination, and long-term reliability risk [6,45]. LECO-style processing offers both opportunity and sensitivity for Cu contacts: it can accelerate desirable interfacial reactions selectively where current density is highest, but any spatial nonuniformity in fluence or barrier integrity increases the probability of transitioning into the uncontrolled regime [45,46].

For Cu-containing contact stacks, the central LECO question is whether localized electrothermal activation improves contact formation without compromising the diffusion-control function of the stack. Compared with Ag-based fired contacts, Cu contacts place greater importance on barrier continuity because local barrier failure can permit rapid Cu transport and uncontrolled silicide growth. LECO may be beneficial when the transient is sufficiently confined to improve interfacial coupling while reducing the need for a higher global firing temperature. Under such conditions, the process can lower contact resistance while limiting wafer-scale thermal exposure. Conversely, if local current crowding or fluence nonuniformity creates electrothermal hot spots at barrier defects, LECO can amplify rather than suppress the risk of Cu penetration, silicide overgrowth, voiding, and leakage. Recent Cu-contacted n-TOPCon results show that optimized firing and LECO can produce screen-printed Cu rear contacts with cell efficiencies above 24%, including a reported 24.3% device with $V_{OC}$ > 730 mV, $J_{SC}$ = 41.1 mA/cm$^2$, and FF = 80.8%, only 0.2% absolute below the Ag-contacted reference [50]. These results indicate that LECO can be compatible with Cu-containing TOPCon contacts, but they also reinforce that barrier integrity and local reaction control must be treated as first-order design variables.

## 3.3 Contact Connectivity and Conduction Pathways

The heterogeneous microstructures described above imply that current transport occurs through a disordered network of microcontacts rather than through a laterally uniform interface [51,52]. A useful abstraction is a random resistor network in which highly conductive nodes (Ag-Si eutectic microcontacts, silicide grains, locally thinned barrier regions) are embedded in a higher-resistance background (glass remnants, oxides, and less-reacted regions) that supports only weak tunneling or field-assisted conduction [53–55]. In such networks, effective conductivity can change sharply once the density and connectivity of low-resistance nodes become sufficient to form robust spanning pathways between the metal and the underlying doped region [56–58]. For LECO-treated contacts, this provides a natural interpretation of threshold-like behavior in $\rho_c$ as a function of laser power, scan speed, or applied bias: once the microcontact population becomes sufficiently connected, the effective transport bottleneck shifts abruptly [57,59,60].

Within this framework, LECO reduces $\rho_c$ primarily by nucleating and enlarging conductive microcontacts at current-crowded sites, thereby increasing both the number of low-impedance injection nodes and their connectivity across the nominal contact footprint.

Below the connectivity threshold, transport is dominated by sparse, non-redundant pathways, so small variations in paste microstructure, glass distribution, or barrier roughness can produce large device-to-device scatter in $\rho_c$. Above threshold, the microcontact network becomes redundant, reducing sensitivity to local variability and improving process tolerance. The same viewpoint clarifies the practical trade-off for process design. Driving the interface deeper into the supercritical connectivity regime improves robustness and lowers $\rho_c$, but excessive reaction increases the probability of over-thickened eutectic or silicide regions, barrier breach, voiding, and passivation damage. Engineering LECO therefore requires controlling microcontact statistics: increasing conductive connectivity density enough to ensure reliable low $\rho_c$ while avoiding the over-reaction modes that compromise recombination, leakage, and long-term stability.

A more quantitative interpretation of this microcontact ensemble can be obtained by separating the nominal metallized area from the electrically active contact area. The key descriptors are the active microcontact fraction, the areal density of low-impedance microcontacts, their characteristic lateral size and depth, and the distribution of local microcontact resistance. In the sub-threshold regime, the active fraction and microcontact density are small, so current is forced through sparse non-redundant pathways and the measured contact resistivity is highly sensitive to local paste, glass, and barrier heterogeneity. During optimal LECO activation, the microcontact density, active area fraction, and lateral connectivity increase together, reducing both the mean contact resistivity and its device-to-device variance. In the over-activated regime, further growth of contact features may continue to reduce apparent contact resistivity, but at the cost of increased metallization-induced recombination, leakage, or local passivation loss. The microstructural optimum is therefore not maximum reaction volume, but maximum electrically useful connectivity at bounded recombination and leakage.

# 4. Electrical Parameter Decomposition Framework

The microstructural pathways summarized in Section 3 are only meaningful to the extent that they measurably change the device electrical behavior. This section introduces a decomposition framework that (i) maps interfacial evolution onto IV-curve signatures, (ii) partitions the measured series resistance into physically interpretable components, (iii) identifies diagnostic indicators of junction damage or shunting, and (iv) consolidates these indicators into a two-dimensional map for interpreting LECO parameter sweeps.

The purpose of this decomposition is to prevent LECO performance from being evaluated using a single metric such as contact resistivity or fill factor. A reduction in contact resistivity is beneficial only if it is achieved without increasing metallization-induced recombination, reducing shunt resistance, or creating stress-sensitive transport pathways. The regime framework therefore treats LECO optimization as a constrained contact-selectivity problem rather than as simple resistance minimization. The desired operating point is the region in which fill factor improves, the pFF-FF gap narrows, open-circuit voltage is retained or improved, and shunt resistance remains stable.

## 4.1 Transport-limited versus junction-limited response

A practical first step is to separate losses that are primarily transport-related from losses that are primarily junction-related [58,59]. Transport losses arise from finite resistivity in the metallization, the emitter sheet, the bulk, and the contact interface [60,61]. They are commonly summarized by an effective series resistance $R_s$ and manifest most clearly as fill factor (FF) loss at approximately unchanged open-circuit voltage $V_{oc}$, provided the junction recombination is not materially altered [60,62,63].

Junction losses arise from recombination and leakage [61,62]. They are reflected in changes to recombination parameters [61,63], as well as in shunt resistance $R_{sh}$ (or equivalently the low-bias leakage slope of the dark IV curve) [64,65]. Junction degradation typically reduces $V_{oc}$, reduces pseudo fill factor (pFF), and modifies the low-voltage curvature and slope of the IV characteristics [62,66].

This distinction is particularly useful for LECO because the intended effect is a reduction of the interfacial transport barrier through the formation and connectivity of conductive microcontacts. The expected electrical response is therefore an improvement in FF with minimal change to junction quality [67,68].

Two electrical signature patterns follow naturally from this split:

1. **Transport-Limited improvement signature**
    - FF increases [68,69].
    - pFF increases or remains high, and the gap (pFF minus FF) narrows [61,66].

- $V_{OC}$ is unchanged within measurement uncertainty, unless deliberate process changes alter junction passivation [68].
- $R_{SH}$ remains stable [64].
- Recombination indicators remain approximately unchanged [61,63].

2. **Junction-limited or damage signature**
   - $V_{OC}$ decreases measurably [61,62].
   - pFF decreases, often independently of any reduction in $R_S$ [63,66].
   - $R_{SH}$ decreases (increased leakage) [64,70].
   - Recombination indicators shift toward recombination dominated behavior [61,62].
   - FF ultimately decreases as junction losses dominate the operating point [62,63].

These patterns from the logical backbone of the regime map are shown in Figure 4.

## 4.2 Contact resistivity and series resistance partitioning

To connect interfacial evolution to IV parameters, the measured series resistance is decomposed into five distinct contributions [71,72],

$$R_s = R_{finger} + R_{emitter} + R_{contact} + R_{bulk} + R_{rear}. \quad (8)$$

Here $R_{finger}$ accounts for resistive loss in the front grid, $R_{emitter}$ for lateral transport in the doped emitter sheet, $R_{contact}$ for the metal-to-contact-stack interface, $R_{bulk}$ for spreading resistance in the wafer, and $R_{rear}$ for rear-side transport through the rear contact stack [72,73].

While this decomposition is formally additive, the terms are not electrically independent in practice [72,74]. LECO primarily targets $R_{contact}$ by reducing the effective specific contact resistivity $\rho_c$ [73,75]. However, lowering $\rho_c$ also weakens current crowding beneath the finger [62,74]. Improved interfacial injection reduces the emitter-side resistive penalty activated under operating current densities, so that the apparent reduction in $R_s$ can include both interfacial and emitter-coupled components [73].

This coupling can be understood through the transfer-width concept. The characteristic length over which current spreads laterally in the emitter before entering the contact decreases as $\rho_c$ is reduced relative to $R_{sheet}$ [62,73]. When LECO lowers $\rho_c$, current transfer becomes less concentrated near the contact perimeter and more uniformly distributed across the contacted region [74]. This redistribution reduces local voltage drop in the emitter adjacent to the finger and suppresses crowding-related dissipation [62]. Consequently, the FF gain associated with a given reduction in $\rho_c$ can exceed what would be predicted by treating $R_{\text{contact}}$ as an isolated lumped element, because part of the emitter-related transport loss is simultaneously relieved [72,74]. Tracking pFF together with FF provides a robust method for attributing observed improvements. pFF approximates the junction-limited fill factor in the absence of series resistance, whereas FF includes transport losses [71,72]. A narrowing of the pFF minus FF gap therefore indicates a transport-dominated improvement even when absolute $R_s$ extraction varies across fitting procedures [72].

In practice, $\rho_c$ is best extracted from dedicated test structures that separate $\rho_c$ from $R_{\text{sheet}}$. Table 1 compiles such extracted values together with device-level metrics [73,75]. Thus, the electrical linkage of the microcontact-ensemble from Figure 3 can be summarized as follows: The microcontact nucleation and coarsening leads to increased microcontact density and connectivity, which reduce mean $\rho_c$, and the reduced $\rho_c$ variability in turn decreases the $R_{\text{contact}}$ and lower the crowding-activated emitter loss, and finally the FF is increased at approximately constant junction quality [72].

*Table 1: Quantitative Electrical Parameters Across LECO Conditions (Data compiled from [3,17])*

| Parameter | Pre-LECO (780°C firing) | Post-LECO (optimal) | Degraded state (bias-T stress) |
|---|---|---|---|
| $\rho_c$, $emitter$ (mΩ · cm²) | 2.9±1.21 | 1.8±0.58 | — |
| $\rho_c$,TOPCon (mΩ · cm²) | 14.1±7.47 | 2.9±0.57 | — |
| $J_o$,e,met at 780°C (fA/cm²) | 761±97 | ~761 (unchanged) | ↑ (spike-induced) |
| $J_o e$,met at 820°C (fA/cm²) | 1656±235 | — (firing T reduced) | — |
| FF (%) | ~80.5 | ~81.8 | ↓ |
| $V_{OC}$ (mV) | ~706 | ~711 | ↓ |
| η (%) | 22.3 (max) | 23.8-24.1 | ↓ |
| $R_C$, front $p^+$-Ag (Ω) | 4.8 | <4.8 | >320 |
| $R_C$, rear $n^+$-Ag (Ω) | 0.7 (LECO-TOPCon) | <0.7 | 392 |
| $R_{SH}$ | Stable | Stable | — |

## 4.3 Shunt resistance and the damage threshold

As laser energy density or applied bias increases beyond the optimal transport-enhancement window, the interface can enter a damaged regime characterized by barrier breach, excessive local reaction, and the formation of parasitic current paths. Although the specific microstructural pathways differ between silver- and copper-based stacks, the electrical signature is common: increased leakage that reduces $R_{\mathrm{sh}}$ and degrades pFF and $V_{\mathrm{oc}}$.

Within this framework, the practical damage threshold is identified electrically. The key indicator is an inflection in $R_{\mathrm{sh}}$ (or an increase in low-bias conductance) that appears once $\rho_c$ and FF gains begin to saturate. The progression is:

- In the transport-limited regime, increasing LECO intensity reduces $\rho_c$ and improves FF while $R_{SH}$ and pFF remain stable.
- Near the boundary, $\rho_c$ reduction and FF gains plateau, indicating that the microcontact network has crossed its connectivity threshold.
- Beyond the boundary, $R_{SH}$ decreases and pFF drops, signaling that additional energy produces junction degradation rather than transport improvement.

Importantly, this boundary can be kinetic. Contacts that appear optimal immediately after processing drift toward damaged-like signatures under subsequent bias-thermal stress, implying that a safety margin from the damaged boundary is required for long-term stability.

## 4.4 LECO Regime Schematics

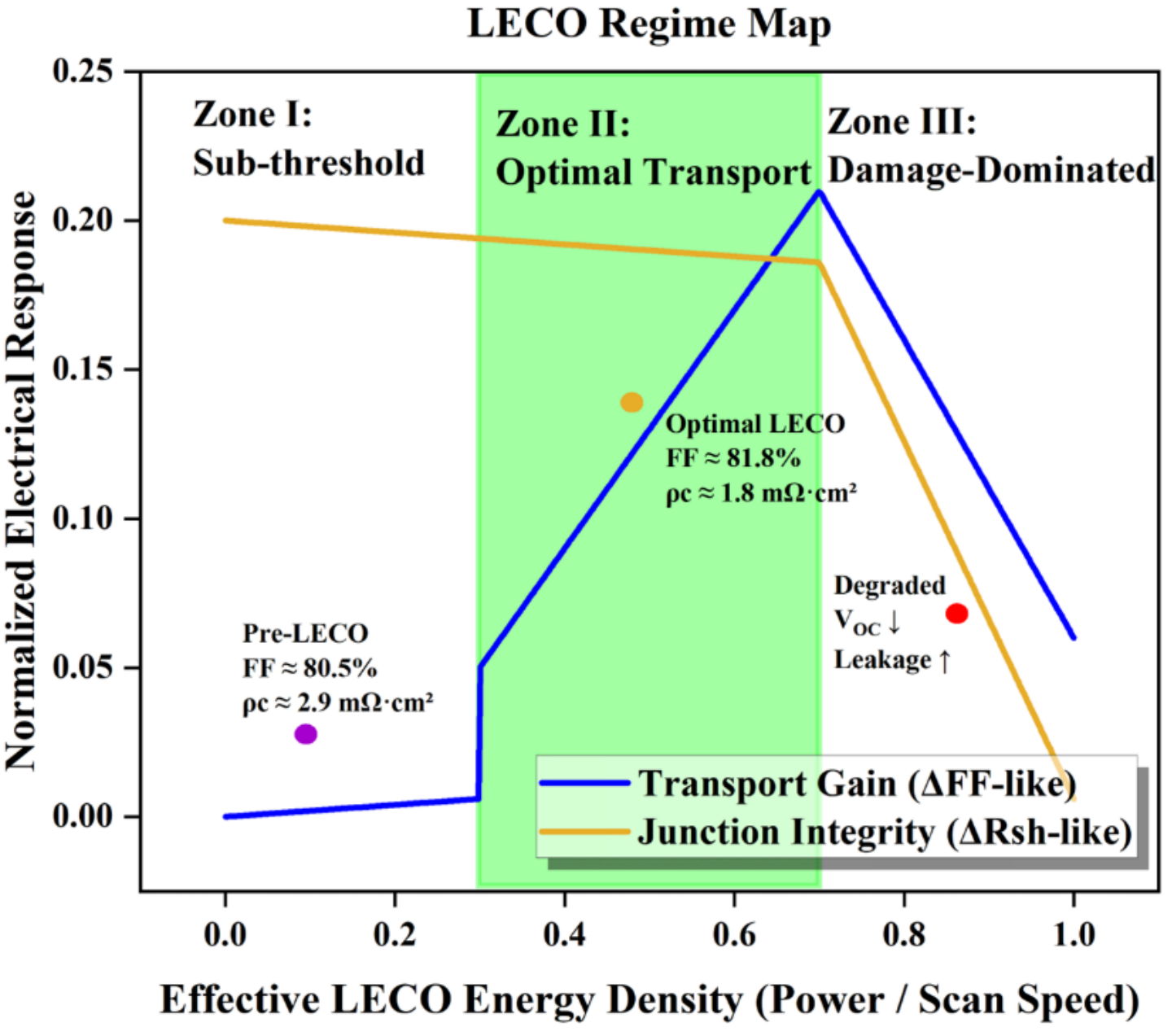

*Figure 4: Conceptual LECO regime map linking effective energy density to electrical response: transport gain (FF-like) increases toward optimal activation, while junction integrity (shunt-like) degrades beyond the damage threshold. Colored regions mark the three diagnostic zones [3,76].*

The diagnostics in section 4.3 can be unified into a two-dimensional schematic with three zones as shown in Figure 4 that convert LECO optimization from trial-and-error into a structured classification task. The x-axis represents effective laser energy density, while the y-axis represents an electrical response metric that combines transport gain with junction integrity. In practice, this can be represented as paired trends, for example the FF-like transport gain alongside an $R_{sh}$-like (leakage) integrity metric, so that improvements in contact transport and the onset of junction degradation can be tracked simultaneously.

The three zones are explained as follows:

**Zone I: Sub-threshold activation.** LECO intensity is insufficient to create a connected population of conductive microcontacts. Electrical parameters remain within measurement scatter.

**Zone II: Optimal transport enhancement.** $\rho_c$ decreases measurably and its variability narrows, leading to a clear FF gain while $V_{oc}$, $R_{sh}$, and pFF remain stable. The width of Zone II defines the practical process margin.

**Zone III: Damaged-dominated regime.** $R_{sh}$ decreases and pFF decreases; $V_{oc}$ declines and recombination indicators worsen. FF may initially remain elevated if $\rho_c$ is very low, but it ultimately decreases as leakage and recombination dominate.

*Table 2: Regime Schematic Diagnostic Signatures (data compiled from [3,17,77,78]).*

| Diagnostic | Zone I: Sub-threshold | Zone II: Optimal transport enhancement | Zone III: Damage-dominated |
|---|---|---|---|
| ΔFF | ≈0 | +1 to +2% abs | Declining, ~50% relative drop under stress |
| $\Delta V_{OC}$ | ≈0 | 0 to + 5 mV (from lower firing T) | Near-constant initially; negative if junction breached |
| $\Delta\rho_c$ | Negligible | >50% reduction, variance narrows dramatically | May appear low initially but unstable under bias-T stress |
| $\Delta R_{SH}$ | Unchanged | Unchanged | Measurable decrease (spike-induced shunts) |
| pFF vs. FF | Both unchanged | pFF↑, FF↑; gap narrows | Both ↓; pFF drops independently of $R_S$ |
| Ideality factor n | Stable (~1.0-1.2) | Stable | Increases (recombination-dominated) |
| $J_{SC}$ | Unchanged | Unchanged or slight improvement | Can collapse (40.116 → 4.33 mA/cm$^2$ under extreme stress) |
| Physical origin | Insufficient thermal/electrical budget for percolation | Low if centered in zone | Spike penetration, junction breached, H-induced barrier formation |
| Reliability risk | None | — | High; H accumulation at $p^+$-Ag interface |

The boundaries between zones shift with paste chemistry, emitter doping and $R_{\text{sheet}}$, passivation stack design, firing conditions, and applied bias. The operational value of the regime schematic lies not in a single numeric threshold, but in a repeatable procedure.

# 5. LECO Across Device Architectures

The LECO drivers developed in Sections 2 to 4, namely optical absorption in the metallization, microsecond-scale thermal transients, and bias-driven current localization, are not architecture-specific. What changes from one architecture concept to another is which layers experience that transient and whether contact activation can be achieved without degrading junction quality or passivation. As a result, the same activation pathway can fall into different regions of the regime schematic depending on thermal tolerance, dielectric integrity, and metallization chemistry. This section evaluates LECO compatibility across PERC, TOPCon, and temperature-sensitive architectures such as HJT and perovskite/silicon tandems using the classification logic introduced in Section 4.4. Table 3 summarizes the architectural parameters that shift the activation and damage boundaries, while Figure 5 provides the common physical basis by illustrating how the transient temperature profile intersects the depth of each architecture's electrically critical interfaces.

*Table 3: Architecture Compatibility Matrix (data compiled from [3,79–82])*

| Parameter | PERC | TOPCon | HJT | [1]Perovskite/Si Tandem |
|---|---|---|---|---|
| Max process T (°C) | 820-850 | 800-900 (poly-Si anneal) | ~200 (a-Si:H limit) | ~130-150 (perovskite limit) |
| Passivation layer at contact | $SiN_x$ ARC | Poly-Si/ 1.2-1.5 nm $SiO_2$ | A-Si:H (5-10 nm) + TCO | TCO + perovskite |
| LECO thermal margin | Moderate | Wide | None (standard LECO) | None (standard LECO) |
| Front $R_{\text{SHEET}}$ (Ω/□) | 100-160 | 200-240 | 50-100 (TCO) | 50-100 (TCO) |
| $\rho_c$ challenge severity | Moderate | High | Low (TCO/metal) | Low (TCO/metal) |
| LECO efficiency gain | +0.37% abs | +0.5-0.6% abs | N/A | N/A |
| Architecture specific risk | BSF under formation at low firing T | Front-side over-activation / local passivation loss | a-Si:H degradation; TCO absorption | Perovskite decomposition; TCO absorption |
| Regime map Zone II width | Narrow (BSF co-constraint) | Wide | Not accessible | Not accessible |

[1] For TOPCon-based perovskite/Si tandems, LECO may be applied to the silicon bottom cell before top-cell integration; therefore, the "not accessible" classification applies to the completed tandem stack rather than to the TOPCon bottom-cell precursor.

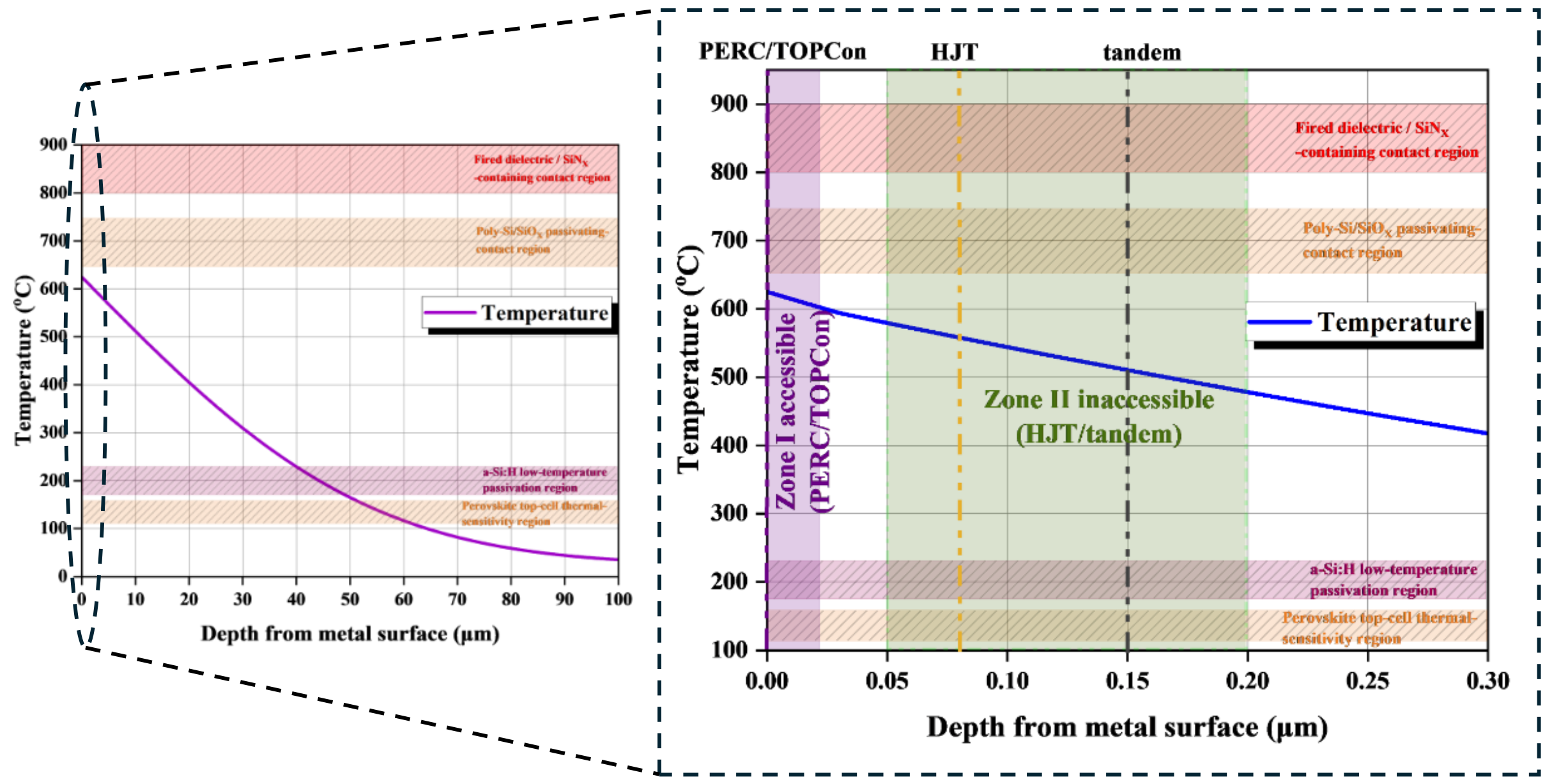

*Figure 5: Schematic LECO transient peak-temperature profile versus depth from the metal surface, shown on device-scale and near-interface scales. Shaded bands indicate qualitative thermal-tolerance regions for fired dielectric or poly-Si-based contacts, hydrogenated amorphous silicon passivation, and perovskite-containing tandem stacks. The figure is intended to illustrate relative thermal-margin differences rather than fixed universal degradation thresholds, because the actual damage boundary depends on dwell time, hydrogen content, layer thickness, capping layers, process history, and local metallization geometry. The zoom indicates that Zone II is accessible for PERC and TOPCon near the contact, but not for HJT cells or completed tandem stacks containing low-temperature passivation or absorber layers [83–91].*

## 5.1 PERC

PERC combines a phosphorus-diffused front emitter with a rear $AlO_x/SiN_x$ passivation stack and localized Al back-surface field (Al-BSF) contacts formed by dielectric opening and Al alloying during firing [92]. LECO was first demonstrated in this architecture and is most naturally interpreted as a method for widening the front-contact process window as the emitter is driven toward lower doping and higher sheet resistance [3,93]. In that operating space, conventional fire-through metallization approaches a contact-formation limit, and LECO provides an additional localized activation channel that recovers FF at reduced global firing severity [3,93]. As summarized in Table 3, reported gains are generally consistent with a shift from Zone I toward Zone II transport enhancement without crossing into junction-limited behavior [3].

Relative to TOPCon front boron emitters, PERC emitters typically operate at lower $R_{sheet}$, which reduces current crowding and relaxes the activation requirement at the metal-semiconductor interface [94]. Consequently, the effective activation threshold is reached at lower LECO intensity, and the marginal benefit can be smaller in standard emitters. The value proposition strengthens as PERC designs move toward ultra-low doping and higher $R_{sheet}$ to reduce recombination and improve spectral response [94,95]. In that regime, LECO functions as a selective contact-activation step that restores transport performance while permitting a lower global firing setpoint, thereby shifting the practical process window toward Zone II [94,96,97].

A key limitation in PERC is that the optimal LECO condition cannot be selected from front-side transport metrics alone because rear passivation and hydrogenation are established during firing [98,99]. Lowering the firing peak to exploit LECO-assisted front contact formation can alter $AlO_x$ activation and redistribute hydrogen released from $SiN_x$, which changes rear-side passivation and, indirectly, $V_{oc}$ stability [98,99]. At the same time, the localized Al-BSF contacts require sufficient thermal budget to form an adequate alloyed region and maintain acceptable rear contact resistivity and recombination [100]. If the firing temperature is reduced too aggressively, rear contact formation becomes limiting even when front contact transport improves [100]. Within the regime-map interpretation, the effective Zone II window in PERC is therefore bounded not only by the front-side damaged threshold, but also by the minimum firing conditions required to preserve rear $AlO_x/SiN_x$ performance and Al-BSF quality [98,100].

## 5.2 TOPCon

TOPCon is the architecture in which LECO has had the largest industrial impact because the contact challenge is more acute and the passivated contact stack can tolerate the transient without catastrophic degradation when processing is kept within Zone II [5]. In a representative TOPCon configuration, the front consists of a boron-diffused emitter, while the rear employs a thin tunnel oxide beneath a doped poly-Si layer [101]. This combination makes TOPCon especially sensitive to front-side contact resistivity and metallization-induced recombination, which explains why LECO has become particularly valuable for this architecture [5,101].

The comparatively wide Zone II window summarized in Table 3 follows from the thermal resilience of the poly-Si stack. The poly-Si layer is formed under high-temperature processing and is generally more tolerant to the brief thermal excursions than amorphous passivation schemes [101]. As indicated by Figure 5, the transient temperature rise decays rapidly with depth, so conditions capable of modifying the metallization interface can remain spatially confined when the process is kept within the optimized regime.

For standard TOPCon cells, the principal LECO design problem is the front contact to the boron-diffused emitter. The front-side must combine low ρc with low metallization-induced recombination, and LECO is attractive because it can improve both quantities without requiring globally aggressive firing conditions [102]. The practical transition from the beneficial regime to the damage-dominated regime is therefore more appropriately associated with excessive local reaction, increased metal penetration, or loss of local passivation near the treated contact rather than with an assumed rear-side tunnel-oxide breach [27,102,103]. When rear-side LECO is applied to n-TOPCon contacts, the poly-Si/$SiO_x$ stack becomes directly relevant, and recent work shows that LECO can reduce Ag/poly-Si contact resistance but may also degrade passivation quality if the process is not properly optimized [104].

In this context, contact activation can be interpreted as a controlled increase in conductive coupling through the doped region via the microcontact ensemble described in Section 3. If reaction volumes become excessive or if the local thermal budget is not adequately bounded, the same mechanism that lowers ρc can increase local recombination near the activated contact and reduce the benefit of passivated-contact design. The objective is therefore to increase microcontact connectivity enough to stabilize transport while preserving local passivation and recombination control within the treated contact region. Figure 5 provides the physical rationale for why TOPCon can maintain this separation more readily than temperature-sensitive architectures.

## 5.3 Advanced and emerging architectures: HJT and Perovskite/silicon tandems

HJT cells and completed perovskite/silicon tandem stacks impose a fundamentally different constraint because their electrically critical passivation and absorber interfaces are intrinsically temperature-sensitive and lie within tens to hundreds of nanometers of the metallization [105,106]. In these completed low-temperature device stacks, the thermal transient that is beneficial for fired contacts can directly degrade passivation quality, shifting the regime schematic such that the damage boundary occurs at or below the activation threshold [105,106].

HJT relies on hydrogenated amorphous silicon layers for passivation, which can degrade irreversibly at relatively low temperatures, producing a rapid loss of interface quality and $V_{oc}$ [107–109]. From Figure 5, the relevant passivation resides extremely close to the heat source, so even modest thermal diffusion from the metal or adjacent conductive layers can exceed the tolerated temperature excursion [108,109]. Under these conditions, the standard LECO mechanism cannot reach Zone II without entering Zone III. Any extension of LECO-like principles to HJT would therefore require an energy-delivery strategy that confines heating to the metal itself or replaces thermal interdiffusion with a non-thermal activation pathway.

Perovskite/silicon tandems, however, should not be treated as a single bottom-cell class. Although tandem research has historically been dominated by silicon heterojunction bottom cells, recent studies have established both conventional TOPCon and bi-poly TOPCon structures as viable bottom-cell platforms for tandem integration [12,20]. In these architectures, LECO can in principle be applied to the TOPCon bottom cell before deposition of the perovskite top cell, allowing the silicon subcell to benefit from reduced contact resistivity and lower metallization-induced recombination without exposing the completed tandem stack to the LECO thermal transient. A recent review of TOPCon-based tandem bottom cells further notes that advances in front-side passivation, doped poly-Si contacts, and laser-assisted firing technologies have narrowed the $V_{OC}$ gap between TOPCon and SHJ cells, strengthening the case for TOPCon as a manufacturable tandem bottom-cell platform [12,20].

After transparent conductive oxides or the perovskite top cell are introduced, the thermal budget tightens further in HJT cells and completed tandem devices [23,110]. At near-infrared wavelengths commonly used in industrial systems, TCOs can exhibit appreciable free-carrier absorption, depositing heat closer to the temperature-sensitive interface than in fired-contact stacks [111,112]. This introduces a secondary heat source near the passivation and further reduces thermal margin [111]. In addition, current flow during any bias-assisted process is mediated by the TCO sheet resistance, which can distribute Joule heating along the finger length rather than localizing it at a metal-to-doped-silicon interface [23,110]. Both effects shift the electrical response toward junction degradation before meaningful contact activation is achieved.

Despite these incompatibilities, emerging architectures stand to benefit disproportionately from reduced metallization usage because their low-temperature pastes typically have higher resistivity and require thicker or wider fingers to meet grid-loss constraints [23]. If an alternative low-temperature activation route could improve conductivity or adhesion of thinner metallization, the material savings would be substantial [113]. For tandems, additional constraints arise from optical shading and the low thermal budgets of the top cell, which further restrict contact formation [113]. For completed HJT-based tandem stacks, additional constraints arise from optical shading, TCO absorption, and the low thermal budget of the top cell, which restrict post-integration contact activation. By contrast, TOPCon-based tandem architectures may exploit LECO at the silicon-bottom-cell stage before top-cell deposition. The relevant constraint is therefore process sequence: standard LECO is generally unsuitable after low-temperature tandem layers are present, but it remains potentially useful upstream for TOPCon bottom-cell optimization.

Overall, the architecture dependence follows directly from the depth and thermal tolerance of the electrically critical interfaces relative to the transient temperature profile in Figure 5. TOPCon offers the broadest practical Zone II margin because it couples a strong contact-resistivity need with a passivation stack that can tolerate brief localized excursions. PERC can benefit, particularly for higher $R_{sheet}$ emitters but is constrained by coupled optimization with rear passivation and Al-BSF formation. HJT cells and completed low-temperature tandem stacks generally reach their damage thresholds before the standard LECO activation window can be accessed, implying that post-integration LECO would require fundamentally different activation strategies.

## 6. Reliability and degradation pathways

The electrothermal conditions that make LECO effective at lowering contact resistivity may also accelerate interfacial evolution under specific stress conditions, although direct evidence for LECO-specific long-term instability remains limited and architecture dependent. In practice, LECO shifts the interface into a highly activated state; the resulting microcontact ensemble and current localization (Sections 3.3 and 4.2) can improve transport, but may also increase susceptibility to diffusion, electromigration, moisture assisted corrosion, and bias thermal instabilities [114]. These risks become more consequential as metallization lines shrink and local current density increases, and as the industry transitions from Ag-based pastes toward Cu-based that rely on diffusion barriers and controlled silicide formation [6]. In this section, degradation mechanisms are linked to their most diagnostic electrical signatures and then organized into reliability classes that complement the instantaneous transport and junction regimes defined in Section 4. For quick orientation, the architecture level susceptibility matrix is summarized in Table 4, while the parameter change rules used to classify stability versus failure under stress are consolidated in Table 5.

The reliability mechanisms discussed below should be interpreted according to their evidence level. Direct LECO-specific evidence is currently strongest for stable PERC module operation and for a limited number of TOPCon bias-thermal instability studies. Evidence for Cu diffusion, electromigration, damp-heat corrosion, and barrier degradation is mechanistically relevant, but in many cases it is drawn from adjacent metallization or module-reliability literature rather than from long-duration LECO-specific datasets. The purpose of this section is therefore to rank plausible degradation routes that define qualification priorities for different metallization stacks and stress environments, rather than to imply that all pathways occur universally after LECO.

## 6.1 Copper diffusion and silicide stability

*Table 4: Reliability evidence relevant to LECO-treated and metallization-modified silicon cell architectures(data compiled from [17,115–119])*

| **Arch/Contact** | **Passivation/Stack** | **Stress (Test)** | **Primary Drift Signature** | **Kinetics (rate or t)** | **Failure Mode** | **Class** |
|---|---|---|---|---|---|---|
| PERC (p-type Ag paste, Al-BSF) | Front: P emitter + ARC; Rear: $AlO_x/SiN_x$ | DH85 (1000 h) | No abnormal changes; passes IEC reliability | - | None reported (no new contact failure) | Stable (S) |
| TOPCon ($p^+$, Ag LECO contact) | Front: $p^+$ Si $+AlO_x/SiN_x$; Rear: $n^+$ Si + $SiO_2$ | 400 oC forward/ rev. bias (0.5 A/0.5 V, 30 min) | RS rises from ~4.8 Ω to > 320 Ω | ~RS↑×60-70 over 30 min (reverse) | H-related passivation at Ag/$p^+$ contact | Latent (L) |
| TOPCon (Cu-plated front grid, non-LECO reference) | Same as above | DH85 (6 h) | $P_{max}$ -11.5% (vs -80% for Ag) | | Dense Cu contact blocking NaCl ingress | Stable (S) |
| TOPCon (p-type, Ag contact + Na contam.) | Same as above | DH85 w/ $NaHCO_3$ (front) | $P_{max}$ -16% in 100 h (from $R_S$↑) | ~100 hr to 16% loss | Ag/Al paste corrodes under ions | Marginal (M) |
| TOPCon (bifacial) | Front: $p^+$ Si w/ ARC; Rear: $n^+$ poly-Si + $AlO_x/SiN_x$ | DH85 (2000 h, white EVA rear) | $P_{max}$ -6% to -16%, primarily $V_{OC}$↓ (rear $J_{02}$ ↑) | 2000 h | Mg in EVA corrodes rear $SiN_x$, raising recombination | Marginal (M) |
| TOPCon (bifacial) | As above | DH85 (2000 h, POE rear) | $P_{max}$ -8% (no unusual failure) | 2000 h | Stable rear passivation | Stable (S) |

For Cu-containing LECO contacts, reliability must be evaluated as a coupled contact-activation and diffusion-barrier problem rather than as a generic Cu-diffusion problem. The same localized energy delivery that can improve contact resistance may also concentrate current and heat at metallization defects, barrier discontinuities, or locally thin interlayers. Therefore, the key reliability question is whether LECO produces a bounded interfacial reaction that remains electrically stable, or whether it creates localized pathways for Cu migration, corrosion-assisted barrier failure, or delayed leakage under damp heat and bias.

Copper is a fast diffuser in crystalline silicon, with transport pathways that include grain boundaries, defective dielectrics, and precipitate networks [120]. If Cu reaches electrically active regions of the device, it can form Cu-rich clusters and $Cu_3Si$, which introduce strong recombination activity and can reduce minority carrier diffusion length [6,121]. For this reason, Cu metallization schemes rely on barrier concepts that both suppress Cu transport and provide a controlled interfacial reaction layer for low resistance contact formation [122].

From a reliability perspective, the critical issue is whether the interfacial reaction remains thin, continuous, and mechanically intact, or whether it transitions into uncontrolled penetration as defined by the microstructural classification in Section 3.2. In the controlled regime, a bounded silicide layer improves adhesion and contact resistivity while blocking further Cu motion [121]. In the uncontrolled regime, excessive thermal budget or local barrier imperfections allow Cu to bypass the intended diffusion block, promoting thicker, irregular silicide growth, voiding driven by unequal atomic fluxes, and local cracking [122,123]. Once cracks or void networks form, they act as fast paths that couple Cu transport to electrically sensitive regions, increasing the probability of deep shunting and junction degradation over time [6]. In LECO assisted Cu concepts, the process window must therefore be defined not only by achieving low $\rho_c$, but also by verifying that silicide thickness, barrier continuity, leakage current, and stress-induced drift remain bounded after accelerated aging.

A second coupling is packaging chemistry. Moisture and encapsulant degradation products can facilitate Cu transport or complexation, effectively lowering the barrier for Cu migration under damp heat and bias [6]. As a result, the same interfacial stack can appear stable or unstable depending on the module environment. For Cu systems, long term stability is therefore a system property of the contact stack plus encapsulation, rather than a metallization only attribute, an interaction that is captured at the architecture level in Table 4.

Recent TOPCon studies show that Cu can also improve contact robustness under specific conditions. Cu plating on screen-printed TOPCon contacts produced a denser front grid and reduced contaminant penetration during damp-heat exposure; in NaCl-assisted DH85 testing, non-plated cells degraded by more than 80% relative after 6 h, whereas Cu-plated cells showed approximately 11.5% relative efficiency loss. This result does not prove universal Cu stability, since the study did not involve LECO-treated Cu contacts, but it shows that Cu-containing metallization can act as a protective contact densification strategy when properly integrated. In contrast, LECO-treated screen-printed Cu-contacted n-TOPCon cells have recently demonstrated efficiencies above 24% and stable mini-module behavior after prolonged damp-heat tests, providing direct evidence that LECO and Cu metallization can be compatible when firing, local activation, and contact-stack design are co-optimized. Taken together, these findings indicate that LECO does not intrinsically worsen Cu reliability, but it makes local barrier quality, contact density, and accelerated-stress validation essential qualification metrics [50,117].

## 6.2 Electromigration under current crowding

Even when diffusion is well controlled by barrier design, LECO activated contacts can remain vulnerable to current density driven transport because current flow at the interface is typically not uniform. As developed in Sections 3.3 and 4.2, conduction can be dominated by a sparse network of high conductance microcontacts and filaments, which increases local current density and concentrates Joule heating at a small subset of sites. Such current localization activates electromigration, especially in fine line metallization where the line cross section is reduced and module operating temperatures can be elevated [6,114].

Electromigration risk is amplified by two reinforcing factors. First, current crowding, which raises the local J and therefore the driving force for atomic flux, while also increasing local $J^2 \rho$ heating, which accelerates mass transport through temperature dependent diffusivities [114]. Second, thermal cycling, which introduces cyclic mechanical stress at interfaces between metal, glass, and silicon [124]. These interfaces are often the same regions where LECO has modified microstructure, redistributed glass, or local residual stresses. Over time, the combined action of electromigration and cyclic stress can nucleate and coalesce voids, thin conductive paths, and promote partial opens in fingers or busbars [114,125].

Electrically, this mode typically appears as a gradual increase in series resistance with a corresponding low FF loss, sometimes punctuated by discrete resistance jumps if a small number of dominant filaments degrade or open [123,124]. In early stages, $V_{oc}$ can remain comparatively stable, consistent with transport limited aging, while variability increases as the current carrying network becomes progressively less redundant, behavior that aligns with the trajectory concept illustrated schematically in Figure 6.

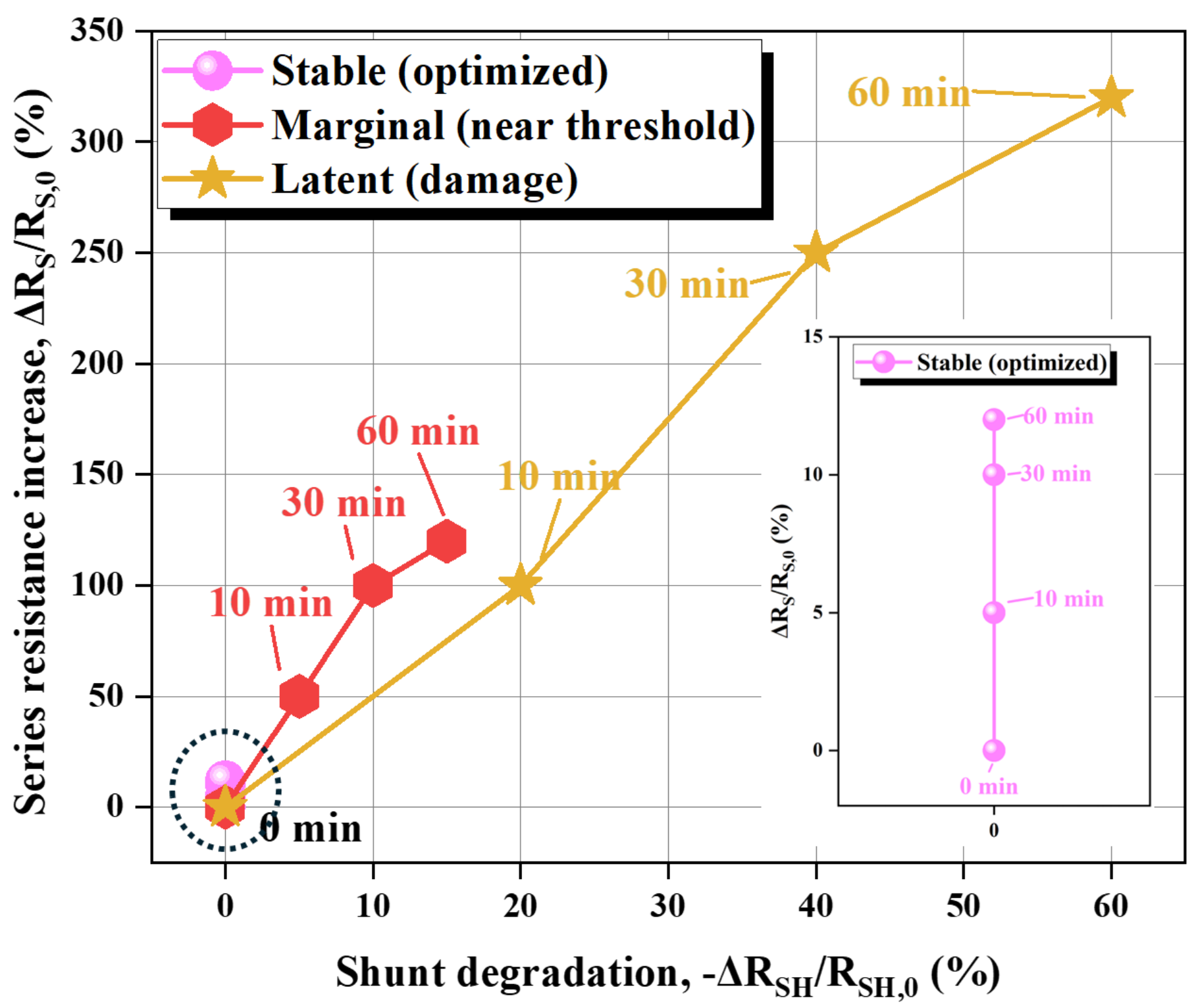

*Figure 6: Schematic $\Delta R_s$ versus $-\Delta R_{sh}$ trajectories illustrating stable (optimized), marginal (near-threshold), and latent (damage) reliability classes under bias-thermal stress. Positive $-\Delta R_{sh}$ denotes a reduction in shunt resistance. Labels indicate stress time (min); the inset zooms the stable regime near the origin [17,115,124,126,127].*

## 6.3 Damp heat corrosion

Environmental exposure introduces chemical drivers that can either act independently (pure contact corrosion) or serve as enablers for other failure modes (barrier degradation that accelerates diffusion). Damp heat exposure accelerates corrosion reactions involving metallization, interfacial glass layers, and encapsulant degradation products [124]. For Ag-based fired contacts, a common pathway is chemical modification of the glass phase at the metal glass silicon interface. Organic acids and ionic species generated during encapsulant aging can leach modifiers from the glass network, weaken adhesion, and create porosity and voids at the triple line region [114,128]. These microstructural changes increase contact resistance and can also promote localized delamination, which further concentrates current and accelerates resistive heating.

Cu-based systems share the same general vulnerability to moisture assisted corrosion but add a diffusion driven failure channel: once corrosion or cracking compromises a barrier, Cu transport into silicon can accelerate, potentially leading to junction leakage and shunt formation [6,122]. In this sense, corrosion and diffusion are coupled rather than independent;

corrosion can be the initiator that turns an otherwise stable diffusion barrier into an active diffusion conduit.

LECO can modify the corrosion landscape in either direction depending on how it restructures the interface. If LECO reduces aggressive glass spikes and improves electrical coupling through a more stable microcontact ensemble, it may lower local overpotentials and reduce the tendency for corrosion driven resistance drift. Conversely, localized thermal and mechanical gradients can introduce microcracks or expose fresh metal silicon protrusions that become preferential pathways for moisture and ionic species [114]. The resulting electrical signatures can therefore differ: corrosion dominated transport loss typically presents as increasing $R_s$ and FF loss with comparatively stable $V_{oc}$ early [124], whereas diffusion assisted damage more often couples transport loss to emerging leakage (declining $R_{sh}$ and increased low bias current), which eventually impacts $V_{oc}$ and pFF [129].

The relative importance of each degradation pathway depends on the contact stack and stress condition. Under dry thermal or bias-thermal stress, resistance drift at the treated contact and hydrogen-related interface changes are expected to be the most LECO-specific risks for Ag-contacted TOPCon. Under damp heat or contaminant exposure, corrosion of the glass-metal-silicon interfacial region and encapsulant-driven ionic transport become more important. For Cu-containing stacks, diffusion-barrier integrity and corrosion-assisted Cu migration are the highest-priority risks, especially when local electrothermal hot spots coincide with barrier discontinuities. Fine-line metallization increases the probability of all current-localization-driven failure modes because fewer microcontacts carry a larger fraction of the current.

## 6.4 Reliability classification proposition

*Table 5: Proposed thresholds for reliability classes (Stable vs Marginal vs Latent) (data compiled from [17,115,116,118,119]).*

| **Parameter (change under stress)** | **Stable (max)** | **Marginal** | **Latent (min)** | **Notes** |
|---|---|---|---|---|
| $\Delta R_S$ (series R increase) | <20% | 20-100% | >100% (doubling) | Latent if Rs >2× initial (e.g. 4.8 Ω in 30 min). Stable typically passes IEC limits (few %) |
| $\Delta R_{SH}$ (shunt R drop) | <20% | 20-50% | >50% drop | Large $R_{SH}$ drop signals leakage. E.g. >100 Ω-cm$^2$ loss often correlates with $V_{OC}$ drop. (Values largely empirical.) |
| $\Delta V_{OC}$ | <10 mV | 10-30 mV | >30 mV | Latent if >~30 mV (~0.5%-$V_{OC}$) is lost; |
| ΔpFF-FF gap | <0.5% abs. | 0.5-1% | >1% | e.g. 16% $P_{max}$ drop in 100 h DH85 corresponded to significant $V_{OC}$ loss. Reduction in pFF gap indicates recombination. |
| Δη ($P_{max}$) | <5% rel. | 5-15% | >15% | IEC61215 implies <5% $P_{max}$ drop (1000 h) is pass. E.g. 16% drop in 100 h is severe |

The regime schematics in Section 4 classifies LECO conditions according to instantaneous electrical response. Reliability introduces a second dimension: whether the interfacial state established in Zone II is kinetically stable under coupled electrical, thermal, and environmental stress [114,124]. In practical terms, the question is not only whether a processing point reaches the low-resistance transport regime, but whether the resulting microcontact ensemble and barrier configuration remain structurally and electrically stable during damp heat, thermal cycling, and bias thermal exposure. Using the parameter change criteria summarized in Table 5, LECO conditions can be organized into three reliability trajectories: (i) stable, (ii) marginal and (iii) latent. These classes are proposed as diagnostic reliability trajectories rather than universal failure modes, allowing LECO-treated contacts to be compared by the magnitude, direction, and time dependence of their electrical drift. These trajectories are visualized in Figure 6; the $\Delta R_{\mathrm{s}}$ versus $-\Delta R_{\mathrm{sh}}$ and in the population-level robustness versus scatter analysis of Figure 7.

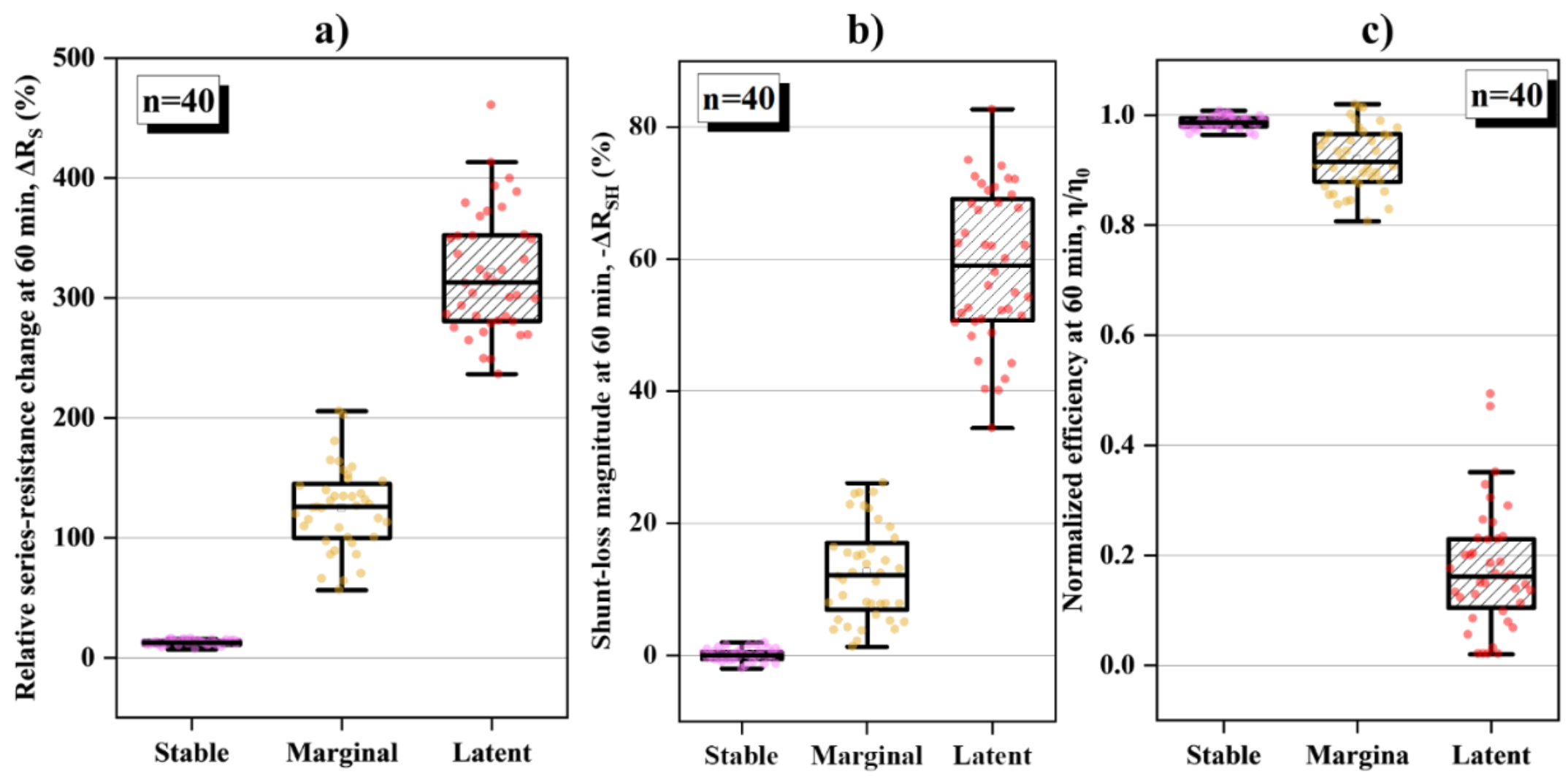

*Figure 7: Schematic population statistics at 60 min stress for LECO reliability classes. Box plots with overlaid data points show (a) $\Delta R_S$(%), (b) $-\Delta R_{SH}$ (%) so larger values mean stronger shunt degradation, and (c) normalized efficiency $\eta/\eta_0$ for Stable (optimized), Marginal (near-threshold), and Laten damage. Stable remains tightly clustered, marginal shows increased scatter, and latent damage exhibits the largest parameter shifts and efficiency loss (data compiled from [5,17,19,79,115,128–132]).*

Processing points that lie well inside Zone II and remain stable during environmental stress exhibit small and approximately monotonic electrical drift. In this regime, degradation is dominated by slow metallization aging rather than abrupt interfacial transitions [123,124]. The fill factor decreases gradually with limited device-to-device scatter, and no systematic leakage paths emerge. The microcontact network remains statistically uniform, current localization is moderate, and diffusion processes proceed at rates consistent with conventional long-term aging rather than instability-driven failure.

As processing approaches either the Zone I to Zone II activation boundary or the Zone II to Zone III damaged boundary, the stability margin narrows. Here the microcontact population may be sparse, or the interface may only marginally exceed the activation threshold. Small variations in glass distribution, barrier continuity, or silicide morphology then produce disproportionately large variations in current localization [6,125]. Under stress, aging accelerates and electrical signatures become mixed: $R_s$ may rise due to contact degradation while localized leakage begins to appear in the weakest regions [129]. The defining characteristic of this regime is high statistical scatter rather than uniform drift, reflecting the sensitivity of transport to local microstructural heterogeneity.

A third trajectory emerges when processing conditions initially place devices within Zone II based on IV metrics, yet the interfacial state is only metastable under prolonged electrical and thermal bias [6,114]. These samples exhibit an incubation period during which electrical parameters remain nearly constant, followed by an abrupt transition. Contact resistance and $R_s$ increase sharply once diffusion, interfacial decohesion, or barrier rupture reaches a critical threshold, while $V_{oc}$ may remain comparatively stable until transport degradation dominates [129]. The time-dependent collapse signature, illustrated in Figure 8 distinguishes this behavior from gradual marginal drift. Because early time $\rho_c$ and fill factor appear acceptable, such metastable states are easily overlooked in short process screens, which motivates explicit stress protocols designed to probe incubation behavior.

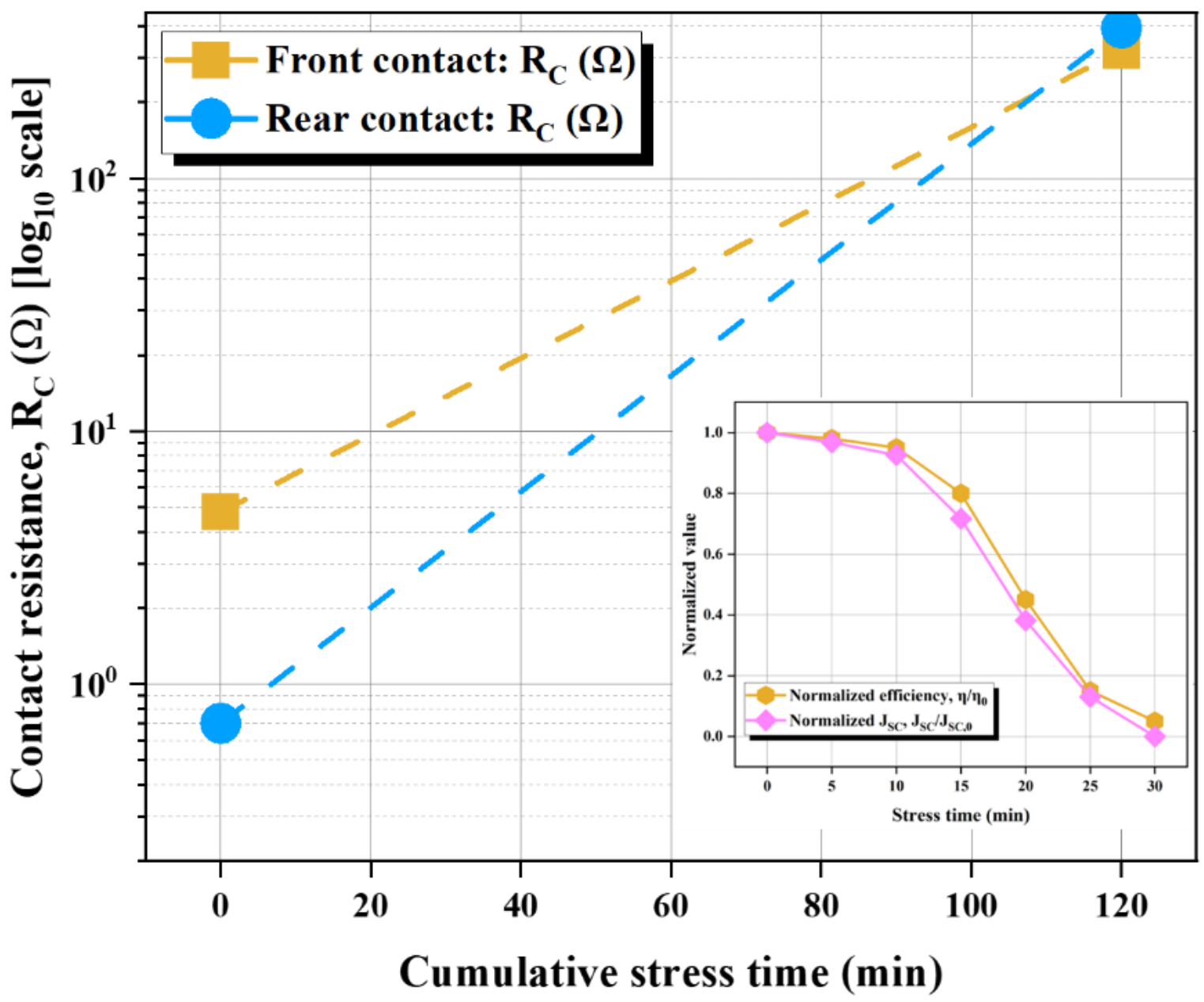

*Figure 8: Incubation-and-collapse signature of latent LECO degradation: contact resistance $R_C$ (front and rear) rises sharply after cumulative stress, with delayed then rapid drops in normalized efficiency and $J_{SC}$ [17,126,127,133,134].*

Framed in this manner, reliability classification becomes a direct kinetic extension of the electrical regime schematics. Deep Zone II processing corresponds to stable optimization, narrow boundary regions correspond to marginal activation with amplified variability, and nominal Zone II points that lack kinetic robustness manifest as latent damage under extended stress [114,124]. The two-axis perspective, instantaneous transport state and long-term kinetic stability, therefore, provides a unified framework for interpreting both initial performance and durability outcomes in LECO-activated contacts.

# 7. Modeling and Predictive Process Design

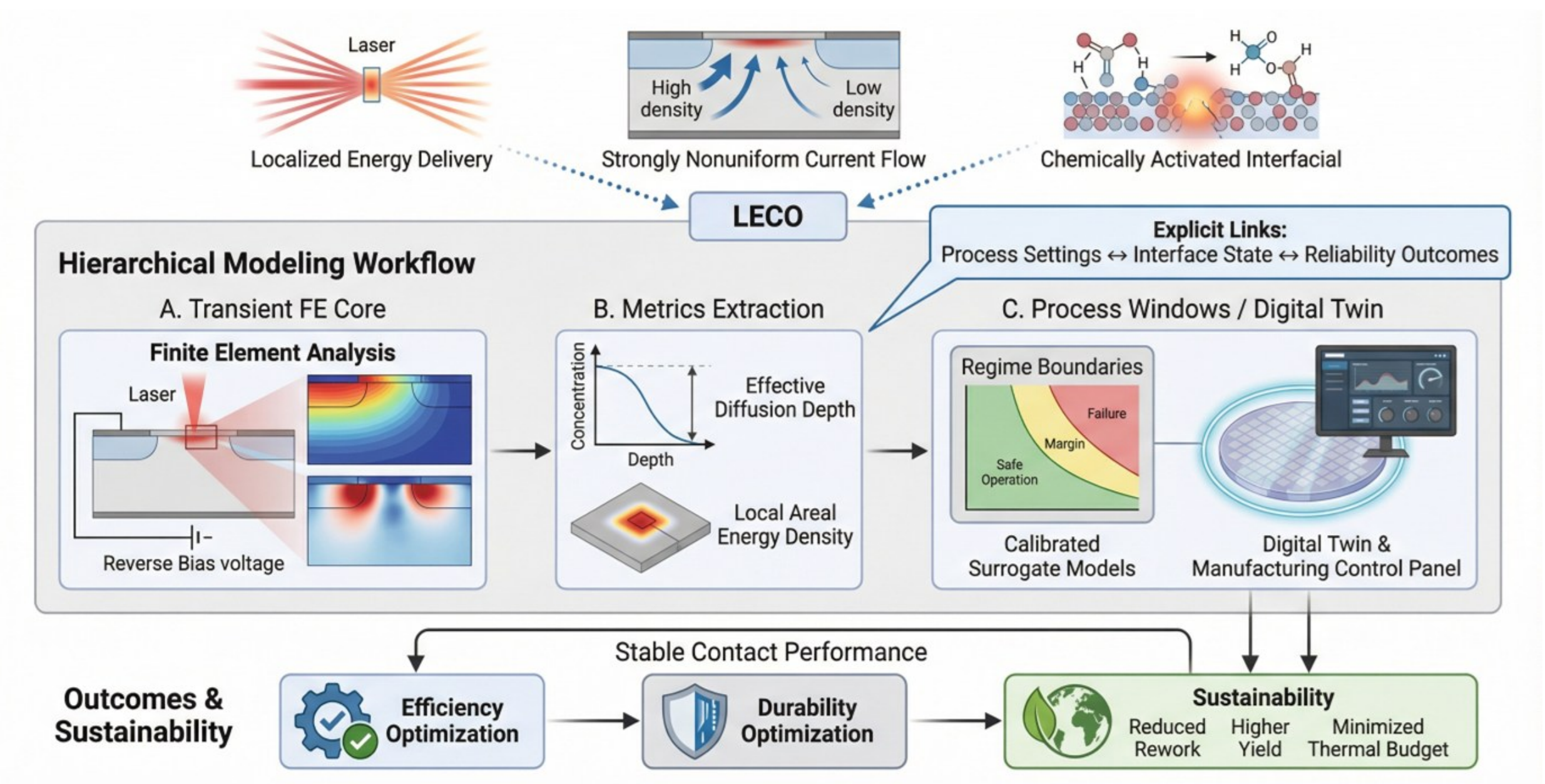

*Figure 9: Hierarchical LECO predictive process-design workflow. A transient electrothermal FE core resolves $T$ and $J$ during laser exposure under reverse bias, these fields are reduced to $L_{\text{eff}}$ and $E_A$, and calibrated surrogate models convert the metrics into process windows and digital-twin-assisted process control linking recipe settings to interface state and reliability outcomes.*

The objective of this section is not to present a stand-alone artificial-intelligence model for LECO, but to define a physics-constrained process-design workflow that can support future data-efficient optimization. The workflow begins with electrothermal field prediction, reduces those fields to physically interpretable state variables, and then uses calibrated thresholds or surrogate models to map recipe space. In this framing, machine-learning tools are useful only when constrained by contact physics and validated against measurable electrical and reliability outcomes. This hierarchy preserves the mechanistic link between transient electrothermal fields, interface-state metrics, and manufacturable control variables, which is essential for translating LECO from empirical recipe tuning to reliability-aware process design.

LECO couples highly localized energy deposition with reverse-bias current localization and thermally activated interfacial evolution, thus, the processed contact state is determined by spatial-temporal fields rather than by a single tool setting. Consequently, purely empirical recipe tuning is sensitive to (i) variability in wafer-to-wafer and tool drift, and (ii) offers limited ability to extrapolate toward long term stability under combined electrical, thermal, and environmental stress. A predictive design framework therefore treats LECO as a coupled multiphysics problem in which process inputs are mapped to interfacial state variables and, in turn, to reliability outcomes, as summarized by the hierarchical workflow in Figure 9. The modeling hierarchy adopted here has three layers that mirror this workflow: (i) a transient finite element (FE) core that resolves the temperature field and current density during laser exposure under reverse bias, (ii) a metric extraction step that compresses these fields into an effective diffusion depth and a local areal energy density, and (iii) a calibration plus surrogate layer that propagates regime boundaries across recipe space to enable manufacturing-oriented process windows and digital-twin control. This hierarchy supports efficiency and durability co-optimization and advances sustainability by reducing rework, yield loss, and excess thermal budget while maintaining stable, low-resistance contact performance [135,136].

## 7.1 Finite element (FE) electrothermal modeling

The foundation is a three dimensional transient FE model that solves heat conduction coupled to electrical current continuity during laser scanning under reverse bias [137,138]. Let $T(x,t)$ be the temperature field and $\phi(x,t)$ the electric potential. The governing equations are

$$\rho(T)c_p(T)\frac{\partial T}{\partial t} = \nabla\cdot(k(T)\nabla T) + Q_{\mathrm{J}}(x,t) + Q_{\mathrm{L}}(x,t) \tag{9}$$

$$\nabla\cdot J = 0, \qquad J = -\sigma(T)\nabla\phi, \tag{10}$$

where $\rho$ is density, $c_p$ is heat capacity, k is thermal conductivity, $\sigma$ is electrical conductivity, and $J$ is current density [137,138]. Joule heating is

$$Q_{\mathrm{J}} = J\cdot E = \sigma(T)\,|\nabla\phi|^2, \qquad E = -\nabla\phi, \tag{11}$$

consistent with standard electrothermal coupling in conductive media [138].

The laser is represented as a moving Gaussian volumetric heat source localized at the metal/Si contact region, which is a common idealization in scanning laser thermal models [139]. Define

$$I_0 \equiv \frac{2P}{\pi r_0^2}, \qquad r^2 \equiv \left(x - x_c(t)\right)^2 + \left(y - y_c(t)\right)^2, \tag{12}$$

where P is laser power, $r_0$ is the $1/e^2$ radius, and $\left(x_c(t), y_c(t)\right)$ is the scan trajectory (for constant scan speed $v$ in $x$, $x_c(t) = x_0 + vt$). A common volumetric form is

$$Q_{\mathrm{L}}(x,t) = \eta_{\mathrm{abs}}\,\alpha\,I_0\,\exp\left(-\frac{2r^2}{r_0^2}\right)\exp(-\alpha z), \tag{13}$$

where $\alpha$ is an effective absorption coefficient in the heated stack and $\eta_{abs}$ accounts for reflectance and coupling. The depth attenuation factor $\exp(-\alpha z)$ follows the Beer Lambert form for absorption in an effective medium [140]. If the interaction is better represented by surface absorption, one may equivalently impose a boundary heat flux $q''_{\mathrm{L}}(r,t)$ with the same Gaussian radial dependence and omit the $\exp(-\alpha z)$ term [139].

Thermal boundary conditions can be written as convection plus radiation on exposed surfaces [137]:

$$-k(T)\nabla T \cdot n = h\,(T - T_{\infty}) + \epsilon\sigma_{\mathrm{SB}}(T^{4} - T_{\infty}^{4}), \tag{14}$$

with heat transfer coefficient $h$, emissivity $\epsilon$, Stefan Boltzmann constant $\sigma_{\mathrm{SB}}$, and ambient temperature $T_{\infty}$. Electrical boundary conditions implement reverse bias via Dirichlet potentials on the terminals:

$$\begin{aligned} \phi &= V_{\mathrm{RB}} \quad \text{on the metallization terminal,} \\ \phi &= 0 \quad \text{on the rear terminal,} \\ J \cdot n &= 0 \quad \text{elsewhere.} \end{aligned} \tag{15}$$

The FE solution provides spatial-temporal descriptors that translate directly to process risk and activation measures. In particular,

$$T_{max} = \max_{x,t} T(x,t)\,, \tag{16}$$

$$G_{max} = max_{x,t}|\nabla T(x,t)|, \tag{17}$$

$$C_J = \frac{max_{x,t}|J(x,t)|}{\langle |J(x,t)| \rangle_{\Omega_{int}}}, \tag{18}$$

where $\langle \cdot \rangle_{\Omega_{\mathrm{int}}}$ denotes an average over an interfacial region $\Omega_{\mathrm{int}}$. These quantities connect to local melting propensity (via $T_{max}$), stress concentration (via $G_{max}$), and current crowding driven instability (via $C_J$). Beyond diagnostics, the electrothermal output provides the primary input to the kinetic layer of the framework, because diffusion and interfacial reactions depend on the temperature history at the contact. The next step therefore maps the predicted interface thermal transient to transport metrics that quantify restructuring intensity.

## 7.2 Diffusion depth and interfacial restructuring

Given the FE predicted temperature history at the interface, interfacial restructuring can be modeled through thermally activated transport of metallic species and, where relevant, dopant redistribution near the junction [141,142]. For a concentration field $c(x,t)$, diffusion is modeled as

$$\frac{\partial c}{\partial t} = \nabla \cdot (D(T)\nabla c), \qquad D(T) = D_0 \exp\left(-\frac{E_a}{k_B T}\right), \tag{19}$$

where $D_0$ is a pre exponential factor, $E_a$ is activation energy, and $k_B$ is Boltzmann's constant [141,142]. When transport is dominated by a short duration event localized near the interface, a one-dimensional reduction yields an effective diffusion depth derived directly from the FE predicted thermal transient:

$$L_{\text{eff}} = \sqrt{2\int_0^{t_{\text{exp}}} D\big(T_{\text{int}}(t)\big)\, dt}, \tag{20}$$

where $T_{\text{int}}(t)$ is a representative interface temperature history and $t_{\text{exp}}$ is the local exposure time. This expression follows from the mean square displacement relation for diffusion with time dependent diffusivity [141]. Separate metrics can be computed for different species, for example $L_{\text{eff}}^{\text{metal}}$ for metal penetration and $L_{\text{eff}}^{\text{dop}}$ for dopant redistribution.

Calibration is performed by fitting $(D_0, E_a)$ or an effective $D(T)$ to depth profiling (SIMS) and cross sectional microscopy, matching either the diffusion front location or the areal inventory of transported species [142]. Once calibrated, $L_{\text{eff}}$ provides a compact measure of restructuring intensity that can be linked to $\rho_c$ reduction while enforcing junction integrity constraints. At the same time, $L_{\text{eff}}$ is inherently a kinetic integral of the thermal transient and it does not explicitly distinguish how the thermal budget is supplied by laser absorption versus reverse bias Joule heating. For manufacturing translation and recipe portability, it is therefore useful to introduce a complementary metric that compresses both energy delivery channels into a single quantity that correlates with activation and damage.

Although this calibration route is physically direct, it is not expected that nanoscale or in situ characterization will be available for every industrial recipe or production lot. A practical implementation should therefore use a tiered validation strategy. High-resolution tools such as TEM, SIMS, atom probe tomography, or operando diffraction are most useful for reference samples, mechanism identification, and threshold anchoring. Routine process control can then rely on macroscopic and inline-compatible proxies, including TLM or CTLM contact-resistivity extraction, Suns-$V_{OC}$, pseudo-FF analysis, dark-IV leakage, EL or PL series-resistance imaging, lock-in thermography, and post-stress drift of $R_S$, $R_{SH}$, and $V_{OC}$. In this hierarchy, nanoscale measurements define the physical meaning of the reduced state variables, while high-throughput electrical and imaging observables provide the practical feedback needed to update process windows.

## 7.3 Local energy density and regime thresholds

A scalar metric that unifies laser input and reverse bias Joule heating is the local areal energy density delivered to the interface:

$$E_A(s) = \int_0^{t_{\exp}} \left(q_{\mathrm{L}}''(s,t) + q_{\mathrm{J}}''(s,t)\right)\, dt, \tag{21}$$

where $s$ parameterizes the interface, $q_{\mathrm{L}}''$ is the laser heat flux at the interface, and $q_{\mathrm{J}}''$ is an interface projected Joule heat flux. In volumetric FE implementation, an equivalent definition integrates volumetric heating within a thin interfacial control volume $V_\delta$ of thickness $\delta$:

$$E_A(s) \approx \frac{1}{A_\delta}\int_0^{t_{\exp}} \int_{V_\delta(s)} (Q_{\mathrm{L}} + Q_{\mathrm{J}})\, dV\, dt, \tag{22}$$

where $A_\delta$ is the projected area of $V_\delta$ on the interface.

With calibrated bounds, $(E_A, T_{max}, L_{\mathrm{eff}})$ fine regimes that separate insufficient activation, stable optimization, and onset of degradation:

$$\begin{aligned}
&\text{No significant modification:}\quad E_A < E_A^{(1)}, \quad \text{or} \quad L_{\mathrm{eff}}^{\mathrm{metal}} < L^{(1)}.\\
&\text{Beneficial optimization:}\quad E_A^{(1)} \le E_A \le E_A^{(2)}, \quad L^{(1)} \le L_{\mathrm{eff}}^{\mathrm{metal}} \le L^{(2)}, \quad T_{max} < T_{\mathrm{crit}}.\\
&Onset of degradation:\quad E_A > E_A^{(2)}, \quad \text{or} \quad L_{\mathrm{eff}}^{\mathrm{metal}} > L^{(2)}, \quad \text{or} \quad T_{max} \ge T_{\mathrm{crit}}.
\end{aligned} \tag{23}$$

Here $E_A^{(1)} and E_A^{(2)}$ are calibrated lower and upper energy density thresholds, $L^{(1)}$ and $L^{(2)}$ are restructuring depth bounds associated with improved $\rho_c$ without junction compromise, and $T_{\mathrm{crit}}$ is a critical temperature associated with melting, barrier rupture, or a sharp increase in recombination risk. These thresholds are identified by fitting model predictions to experimental outcomes, including $\rho_c$, $R_{\mathrm{s}}$, $R_{\mathrm{sh}}$, and stress test drift signatures. Once the regime logic is defined locally, the practical question becomes how to evaluate and visualize it across a high dimensional recipe space. That mapping step motivates surrogate assisted process window construction.

## 7.4 Predictive process window mapping and surrogate modeling

Let the controllable LECO recipe be

$$\theta \in \{P, v, r_0, O, V_{\mathrm{RB}}, \dots\}, \tag{24}$$

where $O$ denotes pulse overlap or an equivalent scan overlap descriptor. The physics derived metric map

$$m(\theta) = \left(E_A(\theta), T_{\max}(\theta), L_{\mathrm{eff}}(\theta), C_J(\theta)\right) \tag{25}$$

is obtained by solving Equations 9 to 20 across representative recipe points. A process window label can then be expressed as

$$y(\theta) = \mathcal{C}(m, (\theta); b), \quad y \in \{0,1,2\}, \tag{26}$$

where $\mathcal{C}(\cdot)$ applies the regime logic in Eq. 23 using the calibrated bound set $b$, and where $y = 0$ denotes no significant modification, $y = 1$ optimized contact, and $y = 2$ risk of degradation.

Because dense sampling by high fidelity FE is computationally expensive, a surrogate $\widehat{m}(\theta)$ is trained on the simulation dataset $\{\theta_i, m_i\}_{i=1}^{N}$ [135]. For Gaussian process regression (one component shown for $E_A$),

$$E_A(\theta) \sim \mathcal{GP}\left(\mu(\theta), k(\theta, \theta')\right), \tag{27}$$

which yields a predictive mean $\widehat{E_A}(\theta)$ and predictive variance $s_{E_A}^2(\theta)$ [135]. The variance enables uncertainty aware windowing, for example

$$\widehat{E_A}(\theta) + \sqrt{\beta}\, s_{E_A}(\theta) \le E_A^{(2)}, \tag{28}$$

where $\beta$ controls risk tolerance in an upper confidence bound style constraint [143]. Experimental observables such as $\rho_c$, IV metrics, and imaging derived proxies can be assimilated by augmenting training targets or by introducing correction factors that map simulated metrics to measure outcomes. The key outcome is that the surrogate provides a fast, uncertainty qualified evaluator of regime membership across the recipe space. This speed is the enabling requirement for digital twin deployment, where predictions must be updated at manufacturing cadence.

## 7.5 Digital Twin formulation and closed loop optimization

With surrogate process windows available for real time evaluation, LECO control can be formulated as a state estimation and constrained optimization problem [136]. Let $x_k$ denote the twin state for run $k$, including effective material and tool parameters such as absorptivity, contact geometry descriptors, and effective diffusion parameters. Let $u_k$ be the applied recipe (a subset of θ) and let $z_k$ be measured outputs (for example $\rho_c, R_s$, and PL or EL features). A generic discrete time model is

$$x_{k+1} = Fx_k + Gu_k + w_k, \tag{29}$$

$$z_k = Hx_k + v_k, \tag{30}$$

where $w_k$ and $v_k$ represent process and measurement noise [136]. The physics and surrogate components provide $F, G$ or, more generally, a nonlinear transition map $x_{k+1} = f(x_k, u_k)$ and measurement map $z_k = h(x_k)$, with inference performed via extended Kalman filtering or Bayesian updates [136].

Recipe selection can be posed as constrained optimization. Define the resistivity residual

$$\Delta\rho_c(\theta) \equiv \widehat{\rho_c}(\theta) - \rho_c^{\star}. \tag{31}$$

Then

$$\min_{\theta} \quad \mathcal{J}(\theta) = w_1\left(\Delta\rho_c(\theta)\right)^2 + w_2\,\mathcal{P}_{deg}(\theta) + w_3\,\mathcal{S}(\theta)$$

$$\text{subject to} \quad \widehat{E_A}(\theta) \le E_A^{(2)}, \quad \widehat{T_{\max}}(\theta) \le T_{\text{crit}}, \quad \widehat{L_{\text{eff}}}(\theta) \le L^{(2)}. \tag{32}$$

Here $\rho_c^{\star}$ is the target contact resistivity, $\mathcal{P}_{deg}$ is a degradation risk proxy derived from proximity to calibrated thresholds or predicted drift, and $\mathcal{S}$ penalizes excessive sensitivity to wafer and paste variability.

In summary, the predictive design framework links process inputs to interfacial outcomes through a compact mechanistic chain: FE electrothermal fields $(T, J)$ define the local thermal budget and current localization, the resulting temperature history drives diffusion and restructuring kinetics through $L_{\text{eff}}$, and the combined laser plus bias contribution is captured by the energetic metric $E_A$ [137,138,141]. Calibrated thresholds on $(E_A, T_{max}, L_{\text{eff}})$ translate these physics into regime boundaries that distinguish insufficient activation, stable optimization, and onset of degradation, enabling explicit control of both performance gain and reliability risk. Surrogate models propagate this regime logic across the high dimensional recipe space with uncertainty awareness, producing actionable process windows at manufacturing cadence [135,143]. Embedded into a digital twin with state estimation, the same hierarchy supports closed loop recipe adjustment under tool drift and material variability, thereby reducing rework and yield loss while constraining thermal budget and interfacial damage pathways [136].

# 8. Research Gaps and Industrial Outlook

LECO has emerged as a practical route to reduce contact resistance, yet broad industrial confidence is still constrained by limited observability of interfacial evolution, incomplete phase-resolved definitions of the processed contact state, and insufficient durability evidence across device architectures and metallization stacks. Long-duration module testing indicates that LECO gains can translate without introducing additional degradation pathways for specific PERC implementations, but recent studies also reveal architecture-dependent instabilities in LECO-treated TOPCon under coupled temperature and bias stress, which elevates bankability concerns and motivates stricter qualification logic [17,115]. The modeling hierarchy in Section 7 offers a physics-informed pathway from process inputs to interfacial metrics and regime boundaries. However, to make those boundaries transferable, the community needs measurements that are time-resolved enough to capture activation and incubation behavior. Plus, chemically specific enough to distinguish competing interface states, and statistically robust enough to quantify variability and long-term drift under module-relevant stress. The resulting research gaps and their translation into near term, midterm, and long-term industrial priorities are summarized in Figure 10. The most pressing gaps cluster into five themes: (i) in situ observation, (ii) nanoscale phase identification, (iii) Cu durability, (iv) fine-line scaling, and (v) data-efficient process optimization.

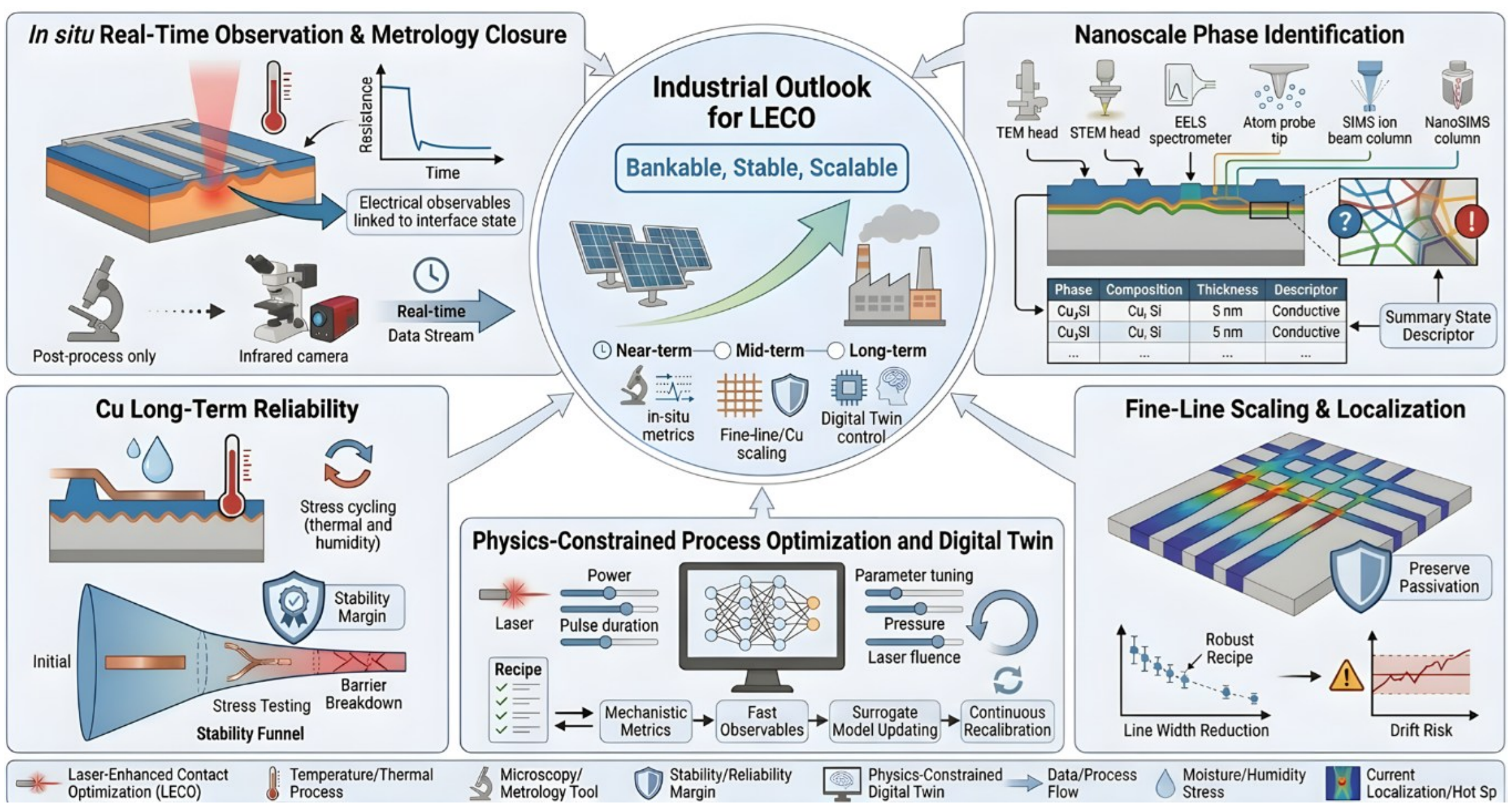

*Figure 10: Research gaps and industrial outlook for LECO. The five near-term to long-term priorities, in situ observables, phase-resolved interface state definition, Cu durability validation, fine-line scaling, and Data-efficient process optimization and digital-twin deployment, converge toward bankable, stable, and scalable deployment.*

## 8.1 In situ real-time observation and metrology closure

A major limitation is the scarcity of in situ diagnostics that follow contact evolution under the coupled conditions unique to LECO, localized laser heating combined with bias-driven current localization. Post-process microscopy can reconstruct end states, but it cannot determine when critical transitions occur, nor can it reliably identify incubation periods that precede abrupt failures. This limitation is particularly acute for passivating-contact architectures, where recent work demonstrates that temperature and bias treatments can drive large series-resistance changes in LECO-treated TOPCon, with spatial nonuniformity evident in luminescence after stressing [17].

Near-term progress is likely to come from pairing electrical observables with spatial imaging and thermal localization. Lock-in thermography is well established for localizing weak dissipative sources and for diagnosing series-resistance related nonuniformities, and it provides a direct bridge from current localization to manufacturable imaging observables [144,145]. In parallel, luminescence-based methods for series-resistance imaging have been demonstrated at throughputs compatible with inline concepts, which supports the use of fast imaging proxies as part of LECO qualification and drift monitoring [146]. A practical target is a minimal observable set that can be mapped onto the Section 7 state metrics, for example proxies for $E_A$ and $C_J$ from thermal localization, and transport-drift signatures from resistance and leakage evolution, then used to distinguish stable optimization from latent damage in real time. Such macroscopic observables do not replace nanoscale phase identification, but they can serve as statistically robust production proxies once their correlation with ρc, $j_{0,met}$, leakage, and stress-induced drift has been calibrated on reference samples.

A second gap is direct access to buried reaction pathways during rapid processing. Operando or in situ diffraction during contact formation has proven decisive for resolving reaction sequences in screen-printed Ag contacts, which illustrates how time-resolved structure evolution can deconvolute competing mechanisms that appear similar in ex situ microscopy [34]. Adapting analogous operando concepts to LECO-relevant stacks, even if only in research settings, would tighten the causal link between process conditions, interfacial state, and subsequent reliability.

## 8.2 Nanoscale phase identification and interface state quantification

Interfacial regions created or modified by LECO can contain multiple nanoscale phases and disordered interlayers whose electrical impact depends on continuity and connectivity rather than average composition. Conventional SEM and EDS often cannot distinguish silicide variants, mixed oxides, and glass-mediated transport pathways at the relevant length scales. This ambiguity becomes limiting because similar values of reduced metrics such as $L_{\text{eff}}$ or $E_A$ can correspond to different interface states, and therefore different stability margins.

Recent LECO-focused microstructural studies underscore the need for a phase-aware state description. Focused-ion-beam enabled cross sections combined with high-resolution electron microscopy have been used to identify LECO-induced microscale and nanoscale contact features at buried Ag/Si interfaces, supporting the premise that a microcontact ensemble can emerge without macroscopic melting [9]. For TOPCon stacks, detailed work on the n-TOPCon rear side shows that current-fired contact formation can differ substantially from non-LECO contacts, and it also highlights that LECO can degrade passivation if not controlled, which makes phase and interlayer continuity central to reliability [130]. Beyond LECO-specific studies, atom probe tomography has demonstrated the ability to resolve hydrogen at the poly-Si/$SiO_x$/c-Si interface in TOPCon, while also clarifying limitations of conventional depth profiling for light elements, which motivates using complementary phase-sensitive methods when the key instabilities are chemistry-driven [147].

A practical community target is a minimum characterization set for LECO contacts that reports (i) phase identity and continuity, (ii) barrier integrity metrics, and (iii) microcontact size and areal density statistics, then links these quantities to electrical regime descriptors and reliability classes. For Cu-containing stacks, depth profiling is additionally required to quantify metal migration pathways, and damp-heat studies already demonstrate that SIMS can detect Cu in the Si bulk after extended stress, which makes diffusion-aware interface descriptors essential for bankability [148].

## 8.3 Cu long-term reliability under LECO-relevant activation

Cu-containing metallization is increasingly attractive for cost reasons, but it introduces durability risks dominated by diffusion, barrier compromise, and moisture-assisted corrosion. These mechanisms are sensitive to local temperature excursions and to current localization, and therefore they interact directly with LECO physics. Broad reviews of Cu-plated metallization emphasize reliability, adhesion, and contact integrity as principal barriers to adoption, with additional constraints arising from outdoor exposure requirements [6]. Earlier reviews of Ni/Cu plating likewise highlight the process complexity and the sensitivity of outcomes to interface preparation and barrier quality, which increases the importance of stability margins when additional activation steps are introduced [149].

The critical evidence gap is long-duration testing of LECO-treated Cu stacks under combined stress profiles that reflect field operation, including temperature, humidity, electrical bias, and thermal cycling. Damp-heat studies of Cu-plated cells demonstrate junction-related degradation consistent with Cu migration, and they show that materials choices such as encapsulant can strongly influence whether Cu transport becomes performance-limiting [150]. Complementary work provides direct evidence of Cu out diffusion through capping layers during damp-heat exposure and detects elevated Cu levels at the Si surface and in the Si bulk, reinforcing the need to treat Cu transport as a bankability constraint rather than a secondary effect [148]. From a module-reliability perspective, electromigration and delamination risks become more consequential as current density increases and as metallization strategies diversify, which further motivates LECO qualification protocols that are incubation-sensitive and spatially resolved [114].

## 8.4 Fine-line metallization scaling and current localization

Fine-line scaling reduces shading and metal consumption, but it increases local current density and narrows the tolerance to interfacial nonuniformity. As finger widths shrink and aspect ratios rise, a small fraction of weak or over-activated contacts can dominate resistance drift and introduce localized heating. Recent fine-line screen-printing work demonstrates 20 μm-class openings and quantifies the coupled impacts on grid geometry and electrical performance, which provides a concrete scaling backdrop for LECO regime transferability [151]. In this scaling limit, robustness margins must be defined in terms of current localization, not only in terms of mean contact resistivity, because geometry-driven current crowding can shift devices closer to the damage boundary even at unchanged nominal recipes.

For passivating-contact architectures, scaling also demands preservation of passivation quality and avoidance of metallization-induced recombination. Reviews of metallization for $SiO_x$/poly-Si passivating contacts emphasize that paste chemistry, thermal budgets, and contact formation pathways remain central to industrialization, which means that any additional local activation step must be evaluated against passivation retention constraints [152]. Interface chemistry further couples into this problem through hydrogen and defect dynamics at the tunnel oxide interface, where nanoscale characterization has shown that light elements and subtle compositional changes can be decisive for stability [147]. LECO-specific evidence on TOPCon stacks shows that passivation degradation can occur when the process is not properly bounded, reinforcing that fine-line scaling and passivation retention are inseparable from the regime-boundary logic in Section 7 [130].

A priority research direction is therefore to establish scaling rules that connect finger geometry and microcontact statistics to allowable ranges of $E_A$, $T_{max}$nd current crowding descriptors such as $C_J$, then validate those rules against long-term drift and scatter. This shifts recipe selection from geometry-specific tuning to transferable design margins.

## 8.5 Data-efficient process optimization and digital twin deployment

LECO is governed by a coupled, high-dimensional parameter space, including laser power, scan strategy, spot size, overlap, reverse bias, upstream firing history, and metallization state. One-factor tuning is inefficient in this setting and increases the likelihood of selecting recipes that sit close to stability boundaries. Data-efficient optimization becomes most valuable when it is constrained by physically interpretable metrics, so that performance improvement is not achieved by moving into regions with elevated reliability risk.

Recent manufacturing-relevant examples show how Bayesian optimization can be integrated into process control problems in silicon PV, including sheet-resistance targeting in doping processes, which is directly analogous to LECO recipe search under constraints [153]. In parallel, digital-twin perspectives in photovoltaics emphasize combining physics-based models with data-driven inference and uncertainty-aware optimization to accelerate innovation while balancing performance, longevity, and sustainability constraints [154]. For LECO, the highest-value industrial implementation is a constraint-aware twin that fuses Section 7 predictors ($E_A$, $L_{\text{eff}}$, $T_{max}$, $C_J$) with fast observables (imaging and electrical tests), then updates margins continuously under tool drift and supply variation.

## 8.6 Industrial outlook

The industrial trajectory of LECO has moved from a corrective post-firing treatment toward a manufacturing-relevant contact-engineering step for high-efficiency TOPCon and PERC technologies. The strongest industrial evidence is now available for TOPCon, where laser-assisted contact formation has enabled simultaneous reductions in contact resistivity and metallization-induced recombination. In Qcells Q.ANTUM NEO cells, LECO produced contacts with $j_{0,met} < 160$ fA/cm$^2$, contact resistivity below 1 mΩ-cm$^2$, and a champion cell efficiency of 25.6% with $V_{OC}$ = 731 mV . The same study reported low degradation of LECO-treated contacts, cells, and modules under damp-heat and thermal-cycling conditions, indicating that optimized LECO can be transferred to production without necessarily introducing a reliability penalty [13].

Jolywood's Special Injected Metallization (JSIM) provides a second industrially important example of laser-assisted firing for TOPCon. JSIM combines customized Ag paste, lower-temperature firing, and whole-surface laser line scanning under reverse bias to improve front-contact formation on industrial G10 TOPCon cells. In damp-heat testing, JSIM cells showed substantially improved stability relative to baseline Ag/Al metallization, with approximately 3.6%$_{\text{rel}}$ PCE degradation compared with approximately 92%$_{\text{rel}}$ for the baseline cells. Glass-backsheet modules using JSIM also showed approximately 2.1%$_{\text{rel}}$ degradation after 1000 h DH85 [116]. These results show that laser-assisted firing can improve not only initial contact quality, but also contact robustness when paste chemistry, firing temperature, and laser activation are co-optimized.

The broader industrial context is that multiple manufacturers are now approaching the voltage regime historically associated with SHJ cells. A recent review of TOPCon-based bottom cells reports that after the development of laser-assisted firing technologies such as LECO, contacted front-side $J_0$ values in mass-production TOPCon decreased to approximately 160 fA/cm$^2$. The same review notes that several manufacturers have reported both-sides-contacted TOPCon solar cells with $V_{OC} > 740$ mV, including JinkoSolar at 741.2 mV [12]. This supports the view that LECO and related laser-assisted firing technologies are part of the industrial pathway by which TOPCon is narrowing the $V_{OC}$ gap with SHJ while retaining compatibility with high-throughput screen-printed metallization and PERC-derived production infrastructure.

Despite this progress, industrial deployment should not be interpreted as a solved reliability problem across all architectures and metallization stacks. Existing results show that LECO-treated PERC modules can be stable, that optimized TOPCon laser-assisted firing can improve damp-heat robustness, and that some TOPCon contacts can still exhibit bias-thermal or damp-heat sensitivity under accelerated stress. The appropriate industrial conclusion is therefore architecture-specific: LECO is already production-relevant for TOPCon and PERC, but bankable deployment requires process windows that jointly control contact resistivity, metallization-induced recombination, leakage, stress-induced drift, and module-level durability. This requirement becomes more important as finger widths decrease and as silver-lean or Cu-containing metallization stacks are introduced, because both trends increase sensitivity to current localization, barrier continuity, and contaminant-assisted degradation.

# Conclusion

LECO has become a practical manufacturing lever for reducing contact resistance in advanced crystalline silicon solar cells, but its outcomes are governed by localized coupling between laser energy deposition, reverse-bias current localization, and thermally activated interfacial evolution. This coupling explains both the existence of an optimization window where $\rho_c$ and FF improve, and the possibility that some LECO-modified interfaces may become metastable under combined electrical, thermal, and environmental stress.

This review unified microstructural and electrical interpretations by treating LECO-modified contacts as heterogeneous microcontact ensembles and mapping their electrical signatures onto a regime framework. The regime schematics (Zones I to III) provides a repeatable method to classify parameter sweeps by instantaneous response, while the reliability regime classification adds a second axis, kinetic stability, to separate stable optimization from marginal activation and latent damage using drift and incubation-and-collapse signatures. This distinction is increasingly critical as metallization scales toward fine lines and as Cu-containing stacks introduce diffusion-barrier and corrosion constraints. By linking instantaneous electrical signatures to time-dependent stability classes, this framework provides a practical basis for separating true process optimization from metastable contact activation.

A hierarchical predictive design approach was outlined to translate these concepts into manufacturable control. A transient electrothermal FE core resolve $(T, J)$, these fields are reduced to $L_{\mathrm{eff}}$ and $E_A$, and calibrated thresholds are propagated across recipe space using uncertainty-aware surrogates suitable for process windows and digital twin control. Closing remaining gaps in (i) in-situ observables, (ii) phase-resolved interface descriptors, (iii) Cu durability validation, and (iv) fine-line scaling rules will determine how broadly LECO can be deployed as a bankable, sustainability-aligned interface activation module.